\documentclass[12pt]{article}
\usepackage{graphicx}
\newcommand{\ccdot}{ \! \cdot \!}
\usepackage{epsfig}

\begin{document}

\title {{\bf  RQM description of PS meson form factors, \\ 
 constraints from space-time translations,\\ and underlying
 dynamics } } 
\author { 
B.  Desplanques$^{1}$\thanks{{\it E-mail address:}  desplanq@lpsc.in2p3.fr},
Y.B. Dong$^{2,3}$\thanks{{\it E-mail address:}  dongyb@mail.ihep.ac.cn}
\\
$^{1}$LPSC, Universit\'e Joseph Fourier Grenoble 1, CNRS/IN2P3, INPG,\\
F-38026 Grenoble Cedex, France \\
$^{2}$Institute of High Energy Physics, Chinese Academy of Science,\\ 
Beijing 100049, P. R. China\\
$^{3}$Theoretical Physics Center for Science Facilities (TPCSF), CAS,\\
Beijing 100049, P. R. China}

\sloppy

\maketitle

\begin{abstract}
The role of  Poincar\'e covariant space-time translations 
is investigated in the case of the pseudoscalar-meson charge form factors. 
It is shown that this role extends beyond the standard energy-momentum
conservation, which is accounted for in all relativistic quantum mechanics
calculations. It implies constraints that have been largely ignored until now 
but should be fulfilled to ensure the full Poincar\'e covariance. The violation
of these constraints, which is more or less important depending on the form of
relativistic quantum mechanics that is employed, points to the validity 
of using a single-particle current, which is generally assumed 
in calculations of form factors. In short, these constraints concern the
relation of the momentum transferred to the constituents to the one 
transferred to the system. How to account for the related constraints, 
as well as restoring the equivalence of different relativistic 
quantum mechanics approaches in estimating form factors, is discussed. 
Some conclusions relative to the underlying dynamics are given in the pion case.
\end{abstract} 


\section{Introduction}
There are many relativistic frameworks in which hadronic systems 
can be described. Each of them has its advantages and its drawbacks. 
Field theory is the most ambitious one but it implies an undetermined, 
possibly, infinite number of degrees of freedom.
Lattice calculations can account in principle for the complexity 
of the QCD interaction but they are limited by their necessary finite size. 
Finally, relativistic quantum mechanics (RQM) offers the advantage 
to rely on a finite number of degrees of freedom. 
It is therefore very close to the simplest view
of the surrounding world, with hadrons considered as a bound state 
of quarks but these ones can only be effective objects.

In this paper, we will be concerned with the application 
of the last framework to the calculation of form factors 
of pseudoscalar mesons (pion and kaon). 
Many approaches were proposed, depending on the symmetry properties 
of the surface on which physics is described \cite{Dirac:1949cp}. 
This entails that some of the generators of the Poincar\'e algebra 
have a dynamical character while the other ones 
have a kinematic one. 
The construction of the algebra was first done 
by Bakamjian and Thomas \cite{Bakamjian:1953kh} and extended later on 
to the front and point forms  \cite{Keister:sb}. 
It relies on the introduction of a mass operator that is Poincar\'e 
invariant and can be used  in any form. 
The different forms were employed for the calculation 
of form factors of various hadrons, mesons as well as baryons. 
In principle, results should not depend on the chosen form \cite{Sokolov:1978}
but one of them may be more efficient than another one. 
Looking at the results of calculations, 
generally based on a single-particle current, 
it was found that they could strongly depend on the form 
for the same solution of the mass operator \cite{Amghar:2002jx} 
or suppose quite different solutions of this operator if they were fitted 
to some measurements \cite{He:2004ba,Julia-Diaz:2004gq,He:2005}. 
This is especially true for the pion charge form factor. 
In such a case, one learns nothing about the underlying dynamics 
of the system under consideration. 
The dependence on the form points to the role of interaction terms 
that are here or there depending on the choice 
of the underlying hypersurface. 

Recently, it was found that constraints stemming from the Poincar\'e
covariance of currents under space-time translations \cite{Lev:1993}
could play an important role in discriminating 
between different approaches for the calculation of form factors 
\cite{Desplanques:2004sp,Desplanques:2008fg}. 
These constraints, go beyond the usual energy-momentum conservation 
which is assumed in all calculations and involve a relation 
between the squared momentum transferred to the system 
and the one transferred to the constituents. 
They imply that the current should necessarily contain 
many-particle terms. The only exception is the case 
of the front form with $q^+=0$. Moreover, an indirect way for accounting 
for these many-particle currents was found with the important result 
that all forms can now produce identical predictions 
for form factors from the same solution of a mass operator. 
However, these results were concerning a pion-like scalar system 
composed of scalar particles, which is not of much interest 
for  physical systems made of quarks that have spin 1/2. 
It is noticed that form factors corresponding to the simplest triangle 
Feynman diagram \cite{Amghar:2002jx} 
could be reproduced exactly \cite{Desplanques:2004sp,Desplanques:2008fg}. 
Those for the Wick-Cutkosky model \cite{Wick:1954,Cutkosky:1954} 
could also be reproduced to a very good accuracy, 
\cite{Desplanques:2004sp,Desplanques:2008fg}, 
pointing in this case to a rather good determination  of the mass operator
\cite{Amghar:2000,Amghar:2000pc,Amghar:2002jx}. 
In both cases, the mass operator has a quadratic form.

In the present work, we want to extend the above work 
for scalar constituents to the case of pseudoscalar mesons 
that are considered as quark-antiquark systems with the goal 
of getting, ultimately, some information on the underlying mass operator. 
Many works have been done for the pion and kaon mesons. 
Though the distinction is not always clear as soon as approximations are made,
they roughly fall in two groups based on field theory 
\cite{Anisovich:1995,Roberts:1996hh,Maris:1998,Maris:2000sk,Merten:2002nz,Simula:2002vm,Bakker:2001pk,deMelo:2002yq,Faessler:2003,Theussl:2004,Bakulev:2004,Noguera:2007,Braguta:2008,Miller:2008,Raha:2009,Bakulev:2009,Dias:2010}
and on relativistic quantum mechanics
\cite{Isgur:1984jm,Chung:1988mu,Cardarelli:1994,Cardarelli:1995dc,Choi:1997iq,deMelo:1997cb,Allen:1998hb,Melikhov:2001zv,Krutov:2002nu,Amghar:2003tx,Coester:2004qu,He:2004ba,He:2005,Li:2008,Desplanques:2009}.
A somewhat different group contains lattice calculations 
\cite{Richards:2004,Alexandrou:2006,Brommel:2007,Simula:2007}.
Within the RQM approach, which we are  interested in here, most works have
relied on the front-form approach with $q^+=0$ but there are also a few works
relying on the instant- or point-form approaches as well as a front-form one
with $q^+ \neq 0$.
Most often, simple wave functions, with a Gaussian 
or a power-law expression, were used. 
For a given wave function, calculated form factors were found to be strongly dependent 
on the form that was used  \cite{Desplanques:2009} and, {\it vice versa}, 
when a fit to measurements was done,  quite different wave functions 
were obtained \cite{He:2004ba,He:2005}. There are only few works 
using a more founded wave function, obtained from a mass operator 
containing both confinement and an instantaneous one-gluon exchange interactions. 
In absence of quark form factors, the first ones 
\cite{Cardarelli:1994,Cardarelli:1995dc}, based on the front form 
with $q^+=0$, tend to overestimate the measurements, 
forcing the authors to introduce some quark form factor 
and, thus, providing some information about this quantity. 
The other one \cite{Desplanques:2009}  was motivated by reproducing 
the asymptotic pion charge form factor in both the front form 
with $q^+=0$ and the Breit-frame instant form. 
The parameters were not especially optimized on some data 
but this work tends also to overestimate data 
though not as much as the first works. Had an other form been used 
(point form, or front form with a parallel momentum configuration, 
or instant form with a large average momentum and also a parallel
momentum configuration), the corresponding form factors 
would underestimate the measurements instead. 
Thus, between the determination of the best form, the role  of the wave function (or the mass operator), 
the contribution of two-particle currents ensuring the right asymptotic behavior, 
the role of possible quark form factors,  we believe 
that there is a sufficiently large number of reasons 
to look at the charge form factor of pseudoscalar mesons in RQM frameworks 
and possible information about the mass operator.
Independently, for a somewhat academic system made of scalar constituents, 
it was found that accounting for the constraints from 
covariant space-time translations was allowing one to recover expressions 
of form factors obtained in a dispersion-relation approach 
\cite{Desplanques:2008fg}. The question arises of whether such a result holds 
for systems consisting of quarks.

The plan of the paper is as follows. In the second section, 
we discuss a number of ingredients relative to the determination 
of the mass operator, which we would like to check ultimately 
from the comparison of form factors calculated using its solutions 
to those actually measured. They concern in particular the linear or
quadratic form of the mass operator, the normalization of the solutions 
and the consequences for form factors in the asymptotic domain. 
The third section is devoted to constraints stemming 
from the Poincar\'e covariance of currents under space-time translations
while their implementation for form factors of pseudoscalar mesons 
is discussed in the fourth section. The role of these constraints 
is illustrated in the fifth section by using a Gaussian wave function 
that is approximately a solution of an interaction containing only confinement. 
Some observations are made in relation with solving some paradoxes 
and restoring fundamental symmetries.  
In the sixth section, we account for one-gluon exchange contributions 
to both the mass operator and the current. We provide values for ingredients 
(string tension, quark mass and QCD coupling) that allow one to approximately
reproduce measurements in the pion case while paying attention 
to the charge radius and the pion decay constant. 
The seventh section contains a discussion of the results and the conclusion.

\section{Mass operator}
\label{mass_op}
The construction of the Poincar\'e algebra in RQM approaches 
supposes that the mass operator fulfills some conditions 
such as independence of the total momentum, of the underlying hyperplane orientation, 
or of the angular momentum but, apart from these general properties, 
not much is known about it. In a two-body system, it depends 
on a 3-dimensional internal variable, denoted $\vec{k}$ here. 
The mass operator therefore looks very much like 
a non-relativistic interaction to which it can be identified in some cases. 
We nevertheless stress that the RQM approach is not a center 
of mass non-relativistic one with some relativistic corrections. 
It is characterized by a deep internal consistency, often ignored, 
which stems from the Bakamjian-Thomas transformation in the instant form 
\cite{Bakamjian:1953kh}
and its generalizations in the other cases \cite{Keister:sb}.

The Bakamjian-Thomas construction \cite{Bakamjian:1953kh}  
was originally involving a linear mass operator, $M=M_0+U $, 
but a quadratic one,  $M^2=M_0^2+ V $, could be used as well 
\cite{Keister:sb}. This choice supposes that the corresponding interaction, 
$V$, be related to the linear-case one, $U$, 
by the relation $V=\{M_0,U\}+ U^2$.
A reason to prefer the quadratic expression is that  $M^2 $ can be identified 
to  $P^2 $, which is more likely the quantity to be considered 
rather than the quantity $ M=\sqrt{P^2} $, but a stronger reason 
is provided by the examination of results 
for the simplest triangle Feynman diagram with scalar constituents
\cite{Amghar:2002jx}. This one represents a theoretical model that is free 
of dynamical uncertainty and, moreover, can be easily studied. 
It provides a minimal set of results that any RQM implementation 
of relativity should reproduce.

Looking at the expression of the charge, which can also be considered 
as a definition of the normalization in the present case, 
it is found that, in the case of unequal-mass constituents, 
it can be written as:
\begin{equation}
F_1(Q^2=0)=\frac{16\pi^2}{N} \!\int\!  \frac{d\vec{k}}{(2\pi)^3}\; 
\frac{(e_{1k}+e_{2k})}{ 2\,e_{1k}\,e_{2k}}\; 
\tilde{\phi}^2(\vec{k}^2)=1 \, ,
\label{eq:norm}
\end{equation}
where the wave function $\tilde{\phi}(\vec{k}^2=k^2)$ is given by: 
\begin{equation}
\tilde{\phi}(\vec{k}^2)
= \frac{1}{(e_{1k}\!+\!e_{2k})^2-M^2} \,, 
\label{eq:wfk2}
\end{equation}
with  $e_{1k} =\sqrt{m_1^2+k^2}$,   $e_{2k} =\sqrt{m_2^2+k^2}$.
The occurrence of the normalization front factor, $16\pi^2/N$, 
may look strange here but it provides some simplification 
of the normalization condition when the integral is performed 
over the Mandelstam variable $s$ rather than the internal momentum variable $k$ 
(see sect. \ref{sec:constraints}). 
While the wave function, $\tilde{\phi}(\vec{k}^2)$, is more appropriate 
for expressing form factors, due to factorizing out typical relativistic 
quantities $1/e_{1k},1/e_{2k}$,  a related expression could be more useful to make 
the relation with the solution of a Hermitian mass operator. The relation is given
by:
\begin{equation}
\phi_0(\vec{k})= \phi_0(|\vec{k}|=k)=\frac{\sqrt{e_{1k}\!+\!e_{2k}}}{ \sqrt{2\,e_{1k}\,e_{2k}}}\;\;
\tilde{\phi}(\vec{k}^2) \,, 
\label{eq:wfk3}
\end{equation}
which corresponds to the normalization condition:
\begin{equation}
F_1(Q^2=0)= \frac{16\pi^2}{N} \!\int\!  \frac{d\vec{k}}{(2\pi)^3}\;\phi_0^2(\vec{k})
= \frac{8}{N} \!\int\! dk \, k^2\;\phi_0^2(k)=1 \, .
\label{eq:norm1}
\end{equation}
The expression of the wave function $\phi_0(\vec{k})$ obtained from 
eq. (\ref{eq:wfk2}) together with eq. (\ref{eq:wfk3}) obviously suggests that it can be 
the solution of an equation with a quadratic mass operator having the form:
\begin{eqnarray}
(M^2-(e_{1k}+e_{2k})^2)\;\phi_0(\vec{k})&=&
-\!\int\!  \frac{d\vec{k}'}{(2\,\pi)^3} \;
\frac{\sqrt{e_{1k}\!+\!e_{2k}}\;\;g_{eff}^2\;\;\sqrt{e_{1k'}\!+\!e_{2k'}}}{
\sqrt{2\,e_{1k}\,e_{2k}}\;\;\;\;\sqrt{2\,e_{1k'}\,e_{2k'}}}\;\phi_0(\vec{k}')\, ,
\label{eq:massop2}
\end{eqnarray}
which, due to its separable form, can be easily solved. 
To make sense, the coupling  $g^2_{eff} $  at the r.h.s. should tend to 0 
so that its product with the diverging  integral 
$\int  d\vec{k}'\,(\sqrt{e_{1k'}\!+\!e_{2k'}} / 
\sqrt{2\,e_{1k'}\,e_{2k'}})\,\phi_0(\vec{k}') $ 
be finite. This does not
prevent one, however, to use the wave function, without any problem, for the
calculation of form factors for the model under consideration.

It is interesting to contrast the above result with what would be obtained 
from a semi-relativistic Schr\"odinger equation in the center of mass:
\begin{eqnarray}
(M-(e_{1k}\!+\!e_{2k}))\;\phi'_0(\vec{k})&=&
-\!\int\!  \frac{d\vec{k}'}{(2\,\pi)^3} \;
\frac{g_{eff}^2}{
2\sqrt{e_{1k}\,e_{2k}}\;\;\;\;2\sqrt{e_{1k'}\,e_{2k'}}}\;\phi'_0(\vec{k}')\,. 
\label{eq:massop1}
\end{eqnarray}
The equation has also a  separable form and, up to a numerical factor, 
its solution is given by: 
\begin{equation}
\phi'_0(\vec{k}) 
=\frac{1}{ 2\sqrt{e_{1k}\,e_{2k}}\;(e_{1k}\!+\!e_{2k}-M)} \,. 
\label{eq:wfk1}
\end{equation}
Comparing this solution with the one for the quadratic mass operator, 
it is seen that they behave differently at large $k$. This feature 
has consequences for the convergence of the norm but, more importantly, 
for estimating the charge form factor at large momentum transfers, 
as far as the  charge form factor extrapolates the expression of the norm 
at $Q^2\neq0$. In the present case, the charge form factor so obtained 
would tend to overshoot the theoretical one, evidencing 
a wrong asymptotic power-law behavior.

The reasons of the above discrepancy have been analyzed in detail, 
what is made possible by the simplicity of the theoretical model 
\cite{Amghar:2002jx,Desplanques:2003nk}.
Actually, considering the center of mass time component of the charge current, 
one should add  to the contribution due to positive-energy constituents: 
\begin{equation}
\Delta J^+=
\frac{16\pi^2}{N}\!\int\! \frac{d\vec{k}}{(2\pi)^3}\;
\frac{ 1}{4e_{1k}\,e_{2k}\,(e_{1k}\!+\!e_{2k}-M)^2}
\,, 
\label{eq:f1p}
\end{equation}
the one due to negative-energy constituents (double Z-type diagram):
\begin{equation}
\Delta J^-=
-\frac{16\pi^2}{N}\!\int\! \frac{d\vec{k}}{(2\pi)^3}\;
\frac{1}{4e_{1k}\,e_{2k}\,(e_{1k}\!+\!e_{2k}+M)^2}\,. 
\label{eq:f1m}
\end{equation}
Considering the sum:
\begin{equation}
\Delta J^{+} + \Delta J^{-}=
2M\;\frac{16\pi^2}{N} \!\int\!  \frac{d\vec{k}}{(2\pi)^3}\;
\frac{ e_{1k}\!+\!e_{2k} }{  
2\,e_{1k}\,e_{2k}\,((e_{1k}\!+\!e_{2k})^2-M^2)^2}=2M\, ,
\label{eq:f1t}
\end{equation}
it is seen that it factorizes into a term that is identical to the norm one
obtained in relation with the quadratic mass operator, eqs.
(\ref{eq:norm}, \ref{eq:wfk2}), and a factor $ 2M $, 
which is nothing but the value of the quantity appearing 
in the time component of the charge current,  $E_i+E_f $, 
calculated in the center of mass. 
Thus, instead of a quadratic mass operator, one could as well use a linear one
but this last choice would require adding further terms in the current so
that to make results consistent with the predictions of the simplest triangle
Feynman diagram. We do not think this is the most efficient way to proceed and
we therefore discard this choice. 

The choice of a quadratic mass operator has further advantages. The invariance
of the charge under boosts is more naturally satisfied with the corresponding
solution, eq. (\ref{eq:wfk2}). In the instant form, for instance, the expression
of the charge for a system with momentum $\vec{P}$ may read:
\begin{equation}
F_1(0)=\frac{16\pi^2}{N} \!\int\!  \frac{d\vec{p}}{(2\pi)^3}\;
\frac{ e_1+e_2 }{ 
2\,e_1\,e_2\,((e_1\!+\!e_2)^2-E^2)^2}\, ,
\label{eq:normP}
\end{equation}
where $e_1=\sqrt{m_1^2+(\vec{P}-\vec{p}\,)^2}$, $e_2=\sqrt{m_2^2+\vec{p}\,^2}$,
$E=\sqrt{M^2+\vec{P}^2}$. In this case, one can easily verify that the
quantities $ d\vec{p}\; (e_1\!+\!e_2)/(2\,e_1\,e_2)$ and  $(e_1\!+\!e_2)^2-E^2$ are
Lorentz invariant, which is not so for the quantities generalizing to an
arbitrary total momentum those given separately by eqs. (\ref{eq:f1p}) 
and (\ref{eq:f1m}). One could add that the suppression of the center of mass 
time component  of the charge current with respect to the scalar one 
by a factor  of the order $M/2m$ is more easily accounted for 
within a framework based on the quadratic mass operator. Actually, 
some of the above properties can be ascribed to a fully relativistic
calculation which involves both forward and backward time-ordered processes,
as given by eqs. (\ref{eq:f1p}) and (\ref{eq:f1m})  for instance, 
often offering convergence properties better than the one with retaining the
first of these processes.

After making the choice of using a quadratic  mass operator, 
the following question concerns the interaction itself. 
In the non-relativistic case, an instantaneous approximation 
is most often used (partly motivated by the success 
of the Coulomb interaction). Again, the consideration of a simple model, 
the Wick-Cutkosky one \cite{Wick:1954,Cutkosky:1954}, is instructive. 
To reproduce the ground-state energy using the instantaneous approximation, 
it was found that the coupling constant should be renormalized significantly. 
This renormalization takes into account contributions that, 
in a time-ordered diagram approach, would correspond 
to retardation effects in a first approximation. 
The charge and scalar form factors calculated in this way were found 
to roughly agree with those calculated from the original Wick-Cutkosky model, 
including the large $Q^2$ domain \cite{Amghar:2002jx,Desplanques:2004sp,
Desplanques:2008fg}. 
A better agreement, 
at the level of a few \%, was obtained by improving the interaction. 
The corrections, of the order of $k^2/e_k^2$ , were chosen so that the
power-law behavior of the interaction in  the high-momentum limit 
be not changed. Interestingly, the interaction in this high-momentum limit 
becomes closer to the bare one. The above results  are important
because they indicate how the interaction entering the mass operator should be
chosen so that to reproduce results expected for form factors in the asymptotic
domain. 

The earlier results that we reminded for the scalar-constituent case 
may be useful for the pseudoscalar mesons we are considering here. 
There is however a significant gap between the two systems. 
The mesons involve spin-1/2 constituents instead of
scalar ones and the exchanged boson has spin 1 instead of 0. 
While, for the  Wick-Cutkosky model, we had to only consider 
the ladder diagram, for a realistic description of mesons, 
we have to consider both the ladder and crossed diagrams. 
Moreover, gluons that are exchanged between quarks carry some color. 
This prevents that the contributions due to crossed diagrams 
and to retardation effects cancel, as they do in the Coulomb case. 
One cannot therefore rely on the successful instantaneous character  
of the Coulomb interaction to justify {\it a priori} the use of a similar
approximation for the gluon-exchange case. 
As a further source of uncertainty, we should
add the confinement interaction and its interference with the gluon-exchange
one. Thus, while the determination of the mass operator in the scalar-particle
case seems to be on a good track, the determination in the QCD case is largely
{\it terra incognita}. The first one can be at most a guide for the second one.
As a starting point, one could consider that the solution of the mass operator,
$\phi_0(k)$, is  described by a Gaussian wave function  for the confining part
and by a term behaving asymptotically like $k^{-7/2}$ (in our conventions
\cite{Desplanques:2009}) for the perturbative part produced 
by a one-gluon exchange. 
We expect that the comparison of theoretical predictions and
measurements will be very instructive to reduce the uncertainty 
on the determination of the pseudoscalar-meson mass operator. 
This supposes that predictions do not depend on the chosen implementation of
relativity however, what we consider in next section.

\section{Constraints from Poincar\'e covariant space-time translations}
It is well known that predictions of form factors in the single-particle current
approximation, when fitted to measurements, lead to solutions of the mass
operator that are strongly dependent on the form that has been used 
\cite{He:2004ba,Julia-Diaz:2004gq,He:2005}. 
As a complementary information, and always in the same approximation 
of a single-particle current, predictions made from the same solution 
of a mass operator exhibit a strong dependence on the form employed 
to implement relativity \cite{Amghar:2002jx}. This situation is quite
unsatisfactory as a correct implementation of relativity should ensure 
that properties of the system under consideration should behave covariantly
under transformations of the Poincar\'e group. These transformation properties 
are somewhat kinematic ones and should not depend on the underlying dynamics. 
As the dependence on a form implies a dependence on the underlying hypersurface on which physics is
described and that changing one hypersurface for another one implies interaction
effects, it can be inferred that
the above calculations of form factors miss some interaction effects. In this
respect, we notice that accounting for constraints related to the covariant
transformations of currents under space-time translations could remove
discrepancies for the scalar system composed of scalar constituents. The above
constraints go beyond  the usual overall energy-momentum conservation that is,
evidently, assumed in all calculations. 
There is no special difficulty in generalizing to spin-1/2 constituents
considered in this work results previously obtained for spin-0 ones. 
We therefore summarize below the main points, providing for some of them a
presentation that complements the one given elsewhere \cite{Desplanques:2008fg}. 

Under Poincar\'e space-time translations, a vector or a scalar current
(denoted $J^{\nu}(x)$ and $S(x)$ respectively) transforms as:  
\begin{eqnarray}
e^{iP \cdot a} \; J^{\nu} (x) \;(S(x)) \; 
e^{-iP \cdot a}=J^{\nu} (x+a) \;(S(x+a)) ,
\hspace*{-2mm} \label{eq:translat1}
\end{eqnarray}
where $P^{\mu}$ represents the 4-momentum operator. 
Making the choice   $a=-x$, one also obtains the relations:
\begin{eqnarray}
&&\hspace*{-2mm}J^{\nu} (x) \;({\rm or}\;S(x))
=e^{iP \cdot x} \;J^{\nu}(0) \;
({\rm or} \;S(0))\; e^{-iP \cdot x} \, .
\hspace*{8mm}\label{eq:translat2}
\end{eqnarray}
Considering the matrix element of the above relations 
between eigenstates of $P^{\mu}$, one obtains the following relation:
\begin{eqnarray}
&&\hspace*{-2mm}<i\;| J^{\nu} (x) \;({\rm or}\;S(x))|\;f>
=e^{i(P_i-P_f) \cdot x}\;<i\;|J^{\nu}(0) \;
({\rm or} \;S(0))|\;f> \, ,
\hspace*{8mm}\label{eq:translat3}
\end{eqnarray}
which allows one to factorize the $x$ dependence.
Combined with the function $e^{iq \cdot x}$ describing the interaction 
with an external probe carrying momentum $q^{\mu}$, and integrating 
over the $x$ variable or assuming space-time translation invariance, one gets
the usual energy-momentum conservation relation:
\begin{eqnarray}
(P_f-P_i)^{\mu} = q^{\mu}\, .
\label{eq:translat4}
\end{eqnarray}
\begin{figure}[tb]
\begin{center}
\includegraphics[width=10cm]{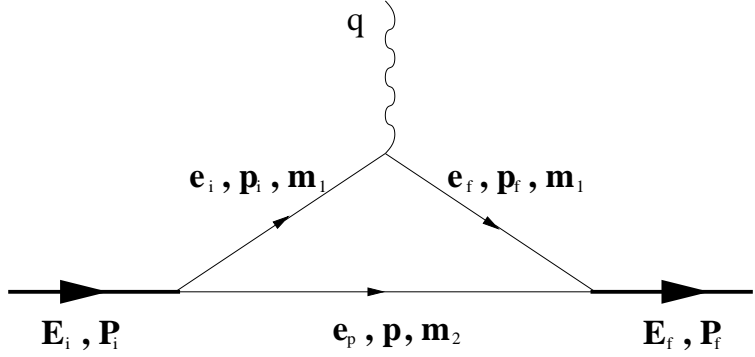}
\end{center}
\caption{Graphical representation of a virtual photon absorption on a
pseudoscalar meson together with kinematic definitions.}\label{fig:1}
\end{figure} 
This relation tells nothing about the relation of the momentum transferred 
to the system, $q^{\mu}$, and to the constituents, $(p_f-p_i)^{\mu}$ 
(see fig. \ref{fig:1} for kinematic definitions). It tells nothing either 
on the single-particle character of the
current, most often assumed, or its many-particle one. This is the place where
further relations involving implicitly the covariant transformation properties 
of the current under translations given by eq. (\ref{eq:translat1}) can be
useful. These ones, mentioned by Lev  \cite{Lev:1993}, involve the commutator 
of the current with the momentum operator $P^{\mu}$ on the one hand, the
derivatives of the current with respect to $x$ on the other hand. 
The simplest one is given by:
\begin{eqnarray}
&&\Big[ P^{\mu}\;,\; J^{\nu}(x)\Big]=
-i\partial^{\mu}\,J^{\nu}(x),
\;\;\; \nonumber \\
&&\Big[ P^{\mu}\;,\; S(x)\Big]=
-i\partial^{\mu}\,S(x) \, . 
\label{eq:translat5}
\end{eqnarray}
At the next order in $P^{\mu}$, a particularly interesting relation is given by:
\begin{eqnarray}
&&\Big[P_{\mu}\;,\Big[ P^{\mu}\;,\; J^{\nu}(x)\Big]\Big]=
-\partial_{\mu}\,\partial^{\mu}\,J^{\nu}(x),
\;\;\; \nonumber \\
&&\Big[P_{\mu}\;,\Big[ P^{\mu}\;,\; S(x)\Big]\Big]=
-\partial_{\mu}\,\partial^{\mu}\,S(x) \, . 
\label{eq:translat6}
\end{eqnarray}
When the matrix element of this relation is taken between eigenstates 
of $P^{\mu}$ and assuming a single-particle current, one gets, 
after factorizing the $x$ dependence as done in eq.~(\ref{eq:translat3}):
%
\begin{eqnarray}
&&<\;|q^2\; J^{\nu}(0) \;({\rm or}\;S(0))|\;>
=
<\;|(p_i-p_f)^2\,J^{\nu}(0)\;({\rm or}\;S(0))|\;> \,,\hspace*{8mm} 
\label{eq:translat7}
\end{eqnarray}
where $q^2$ represents the squared momentum transferred to the system 
and $(p_i-p_f)^2$ the one transferred to the constituents. 
We observe that the relation could be satisfied in a field-theory approach,
where $(p_f-p_i)^{\mu} = q^{\mu}$, but, until recently, it was not checked 
within RQM approaches where it turns out to be generally violated.
In this case, the violation shows that the assumption of a single-particle
current is not supported by the above constraints. The current should then
involve many-particle terms and it can be hoped that these ones contribute to
restore the equivalence of different forms for the calculation of form factors 
\cite{Sokolov:1978}.
Interestingly, it is found that eq. (\ref{eq:translat7}) is fulfilled 
in the front-form approach with $q^+=0$ (see below), providing in this case 
support for the assumption of a single-particle current. In all the other cases, 
the current should involve many-particle terms to satisfy 
the covariant character of translations evidenced by eq. (\ref{eq:translat1}). 
Calculating the contribution of many-particle terms is quite tedious 
and this has been done for a limited number of cases with the aim 
of accounting for current conservation \cite{Desplanques:2003nk}
or getting the expected asymptotic behavior of the charge form factors 
for the pion \cite{Desplanques:2009} or for a system of scalar constituents 
in the point form-approach \cite{Desplanques:2003nk}. 
Moreover, if these extra terms restore the equivalence of predictions 
of different approaches, we expect that they should occur at all orders 
in the interaction. However, the fact that the form factor in the front form 
with $q^+=0$ satisfies eq. (\ref{eq:translat7}) with a single-particle current 
suggests that the task is not hopeless and that some simplification could occur.

\section{Form factors with implementation of constraints}
\label{sec:constraints}
In this section, we provide information about various quantities that will be
calculated in the next sections. They often represent a generalization 
of expressions given elsewhere for equal-mass constituents. This is also done
with the aim to facilitate the comparison with other works, as far as notations
or conventions are concerned.

We first remind the definitions that we are using for the charge 
and scalar form factors, $F_1(Q^2)$ and $F_0(Q^2)$ respectively, 
in the case of an interaction of constituent number 1 with the external probe:
\begin{eqnarray}
&&\sqrt{2E_i\;2E_f} \;\langle \;i\;|J^{(1)\mu}_{op.}|\;f\; \rangle
= (P_i+P_f)^{\mu} \;F^{(1)}_1(Q^2)\,,
\nonumber \\
&&\sqrt{2E_i\;2E_f} \;\langle \;i\;|S^{(1)}_{op.}|\;f\; \rangle
= 4m_1 \;F^{(1)}_0(Q^2)\,.
\label{defs_ff}
\end{eqnarray}
The single-particle part of the quark operators $J^{(1)\mu}_{op.}$ 
and $S^{(1)}_{op.}$, we give here for fixing some relative overall factors, 
are given by $\bar{q} \gamma^{\mu} q $ and $\bar{q}q$ respectively. 
They determine the strength of possible many-particle contributions 
that should be considered, such as Z-type-diagram ones for instance. 
These contributions, which depend on the approach under consideration, 
have an interaction character.

To account for the  constraints mentioned in the previous section, 
we proceed as follows.\\
{\it i}) We assume that the current keeps the structure 
of a single-particle current otherwise it would be difficult 
to recover the front-form results with $q^+=0$.\\
{\it ii}) The discrepancy between different approaches involves interaction
effects that are here or there depending on its choice. 
Looking at the expression of wave functions entering calculations, 
it is found that they differ by interaction effects that are located 
in the coefficient of the momentum transfer, $q$, or some related quantity. 
Our proposal is to multiply this quantity by a factor $\alpha$  
so that to fulfill eq. (\ref{eq:translat7}) and incorporate in this way 
interaction effects that have been missed. 
We thus obtain the equation:
\begin{eqnarray}
&&\hspace*{-10mm}q^2=
``[(P_i\!-\!P_f)^2
+2\, (\Delta_i \!-\! \Delta_f)\;  (P_i\!-\!P_f) \cdot \xi
+(\Delta_i \!-\!\Delta_f)^2 \;\xi^2]"
\nonumber \\
&&\hspace*{-6mm}=\alpha^2q^2-2\,\alpha ``(\Delta_i \!-\!\Delta_f)" \;q \cdot \xi
 +``(\Delta_i \!-\!\Delta_f)^2"\;\xi^2 \,,
\label{eq:q2}
\end{eqnarray}
where the quantity $\Delta$ holds for an interaction effect, of which expression 
given in ref. \cite{Desplanques:2008fg} is reminded here:
\begin{equation}
\Delta=\frac{s-M^2}{
\sqrt{ (  P \ccdot \xi)^2 + (s\!-\!M^2)\;\xi^2 } 
+   P \ccdot \xi }\,. 
\label{Delta}
\end{equation}
The 4-vector $\xi^{\mu}$ represents the orientation of the hyperplane 
on which physics is described.
It appears in the equation that relates the momenta of the constituents, the
total momentum of the system, and the quantity  $\Delta$  
\cite{Desplanques:2008fg}: 
\begin{equation}
(p_1+p_2)^{\mu}=P^{\mu}
+  \Delta \; \xi^{\mu}  \, .
\label{sump1p2} 
\end{equation}
Expressions of $\xi^{\mu}$  for different approaches may be found in ref.
\cite{Desplanques:2008fg}\footnote{In the case of the ``instant form with the
symmetry of the point form" \cite{Bakamjian:1961} that has been used in many
works, two 4-vectors $\xi_i^{\mu}$ and $\xi_f^{\mu}$ describing the velocity of the
initial and final states must be introduced. 
In this approach, which is denoted here ``P.F." and is different 
from the point form proposed by Dirac \cite{Sokolov:1985jv}, 
the above equation becomes slightly more complicated but is still manageable.}.
The notation $`` \cdots "$ in eq. (\ref{eq:q2}) reminds that the corresponding quantity should
include the effect of the constraints. It is immediately seen that, 
in the front-form case where $\xi^{\mu}$ is often
denoted $\omega^{\mu}$ with $\omega^2=0$,  this equation 
is satisfied  for the momentum configuration
$q^+=\omega.q=0$ with $\alpha=1$. In this case, the equality 
of the squared momentum transferred to the system 
and to the constituents, eq. (\ref{eq:translat7}),  is trivially fulfilled. 
In the other cases, one has to take into account the modification 
of the calculation given by the coefficient $\alpha$, 
which is solution of eq. (\ref{eq:q2}). Consistently with our intent, the
coefficient $\alpha$ departs from the value 1 by interaction effects.
While the coefficient $\alpha$ was given an approximate numerical value 
in earlier works, we stress that it is given here its exact algebraic expression 
as obtained from solving the above equation, similarly to what has already 
been done in ref. \cite{Desplanques:2008fg} for scalar constituents.

The practical implementation of the constraints discussed above for the form
factors of pseudoscalar mesons does not differ much from the one for a scalar
system composed of scalar constituents \cite{Desplanques:2008fg}. 
As for this system, there are two form factors corresponding 
to Lorentz 4-vector and scalar currents, $F_1(Q^2)$ and $F_0(Q^2)$ 
respectively, which sum up the contributions of constituents 1 and 2 with the
corresponding charges. 
The main change concerns the introduction of the quark-spinor description. 
In short, this change amounts to multiply the integrand for scalar constituents, 
mostly given in ref. \cite{Desplanques:2008fg}, 
by the ratio of the matrix elements of the free-particle currents for spin-1/2 
and spin-0 constituents.
The most general expressions of form factors for pseudoscalar mesons 
considered here are somewhat cumbersome. They involve results 
for different forms, without and with the effect of the implementation 
of the constraints and for different quark masses.
Due to their technical character, we refer to a separate note 
\cite{Desplanques:2009un} for details. In this note,  it is also shown how
charge form factors in different approaches, after implementing the effect of
constraints, can be identified to a unique expression, which turns out to be 
the one obtained from a dispersion-relation approach. 

First examples of a relationship between this last approach and a RQM one 
were mentioned for the front-form case with $q^+=0$ 
\cite{Anisovich:1992hz,Melikhov:2001zv}. Quite generally, 
such identities are non-trivial. On the one hand, the dispersion approach 
implies a two-dimensional integration over Mandelstam variables 
of the covariant interaction-free scattering amplitude of the constituents 
with the external field. As a result, the squared momentum transferred to the
constituents, $(p_i-p_f)^2$, and to the system, $q^2$, are equal 
and eq. (\ref{eq:translat7}) is automatically fulfilled. 
On the other hand, the RQM approaches imply generally a
three-dimensional integration over the momentum of one of the constituents. 
Different choices are possible for the matrix element of the current 
which is integrated over in this case (see ref. \cite{Amghar:2002jx} 
for an example). Matrix elements used in the present work are consistent 
with what is expected in the interaction-free case. They nevertheless contain
extra terms that have an interaction character and complete those accounted 
for by the factor $\alpha$ discussed above. These terms ensure that the form factor
at $Q^2=0$, which is not affected by the above constraints, should be Lorentz
invariant. This is illustrated in a particular case by the consideration 
of eqs. (\ref{eq:f1p}), (\ref{eq:f1m}) and  (\ref{eq:f1t}) of sect. \ref{mass_op}.

\subsection{Expression of form factors in the dispersion-relation approach}
As the dispersion-relation approach is the one which the other RQM approaches
are expected to converge to after implementing the effect of constraints, 
we give here the corresponding results. Though we are mainly interested here 
in the charge form factor $F_1(Q^2)$, we also include 
the scalar form factor $F_0(Q^2)$ as the comparison of the two form factors 
shows features significantly different from the scalar constituent case, 
especially with respect to their asymptotic behavior. 
The expressions for the contributions of the constituent 1 read: 
\begin{eqnarray} 
&&F^{(1)}_1(Q^2) =\frac{1}{N} \int d\bar{s} \;  d(\frac{s_i\!-\!s_f}{Q}) \; 
 \phi(s_i) \; \phi(s_f)\;
 \nonumber \\
&& \hspace*{2cm}\times\frac{\Big[ 2s_i\,s_f\!-\!\Delta m^2(2\bar{s}  \!+\!Q^2)
\!-\!(m_1\!-\!m_2)^2(2\bar{s}\!-\!2 \Delta m^2 \!+\!Q^2) \Big]
\;\theta(\cdots) }{
D\sqrt{D}\sqrt{s_i-(m_1\!-\!m_2)^2}\;\sqrt{s_f-(m_1\!-\!m_2)^2}} \, ,
\nonumber \\
&&F^{(1)}_0(Q^2) = \frac{1}{N} \int d\bar{s} \;  d(\frac{s_i\!-\!s_f}{Q})\;
 \phi(s_i) \; \phi(s_f) 
\nonumber \\ 
&&\hspace*{2cm} \times \frac{\Big[2m_1
\Big(\bar{s}\!-\!(m_1\!-\!m_2)^2\Big)+m_2\,Q^2\Big] \;\theta(\cdots) }{
2\sqrt{D}\,(2m_1)\,\sqrt{s_i-(m_1\!-\!m_2)^2}\;\sqrt{s_f-(m_1\!-\!m_2)^2} } \, ,
\label{eq:dp15}
\end{eqnarray}
where $m_1$ refers to the interacting constituent, $m_2$ refers to the
spectator one and $\Delta m^2=m_2^2-m_1^2$. The quantities $\bar{s}$, 
$D$ and $\theta(\cdots)$ are defined as:
\begin{eqnarray}
&& \bar{s}=\frac{s_i+s_f}{2}\,,
\nonumber \\ 
&& D= 4\bar{s}+Q^2+\frac{(s_i\! -\! s_f)^2}{Q^2}\, ,
\nonumber \\ 
&& \theta(\cdots)=\theta\Big(\frac{s_is_f}{D}\,c_{\Delta m^2}
-m_2^2\Big)\,,
\nonumber \\ 
&&{\rm with}\;\; c_{\Delta m^2}=\Big(1\! +\! \frac{\Delta m^2}{s_i}\Big) \Big(1\! +\! \frac{\Delta m^2}{s_f}\Big)
+\frac{\Delta m^2(s_i\! -\! s_f)^2}{Q^2s_is_f}\, .
\label{eq:dp16}
\end{eqnarray}
In the equal-mass constituent case, where there is  no difference between
contributions of constituents 1 and 2, the above expressions simplify to read:
\begin{eqnarray}
&&F_1(Q^2) =\frac{1}{N} \!\int\! d\bar{s} \;  d(\frac{s_i\!-\!s_f}{Q}) \; 
 \phi(s_i) \; \phi(s_f)\;
\frac{ 2\sqrt{s_i\,s_f}
\;\theta(\cdots) }{D\sqrt{D}} \, ,
\nonumber \\
&&F_0(Q^2) =\frac{1}{N} \!\int\! d\bar{s} \;  d(\frac{s_i\!-\!s_f}{Q}) \; 
 \phi(s_i) \; \phi(s_f)\;
\frac{ (2\bar{s}+Q^2)
\;\theta(\cdots) }{ 4\sqrt{s_i\,s_f}\sqrt{D}} \, .
\label{eq:disp}
\end{eqnarray}
It is noticed that the above form factors differ from the ones for scalar
constituents by making the exchange of factors $2\sqrt{s_i\,s_f}$ and 
$2\bar{s}+Q^2$ at the numerator. This immediately shows that the charge form
factor will decrease asymptotically faster than the scalar one in the pion case
compared to the scalar-constituent case. In this case, the ratio is given by a
factor 2 which represents the large $Q^2$ limit 
of the ratio $2(2\bar{s}+Q^2)/D$.

We notice that the above expression for the charge form factor agrees 
with the one given by Melikhov  \cite{Melikhov:2001zv} 
but disagrees with an other one given in ref. \cite{Krutov:2002nu} 
for equal-mass constituents.
The discrepancy factor in the integrand, $(s_i+s_f+Q^2)/(2 \sqrt{s_i\;s_f})$, 
is the same as the factor found for scalar constituents \cite{Desplanques:2008fg}.
In this case, expressions of form factors 
were checked by considering the simplest Feynman 
triangle diagram,  including unequal constituent masses or different masses for 
the initial and final states. In the pion case, the comparison supposes to
disentangle the effect of the Wigner rotation used in one 
of the approach \cite{Krutov:2002nu} (see appendix \ref{app:A}).

\subsection{Expression of form factors in the front-form approach 
with $q^+=0$}
As expressions of form factors in the front-form case with  $q^+=0$ are not
affected by constraints related to space-time translations, we also give their
expressions. They read:
\begin{eqnarray} 
F^{(1)}_1(Q^2) & \!=\! & \frac{1}{\pi\,N} 
\int \! d^2R \int_0^1\! \frac{dx}{x(1\!-\!x)} \; 
\frac{I^0_{\omega}}{\tilde{I}^0_{\omega}} \;
\phi(s_i) \; \phi(s_f)   \;,
\nonumber \\ 
F^{(1)}_0(Q^2) &\! =\! & \frac{1}{\pi\,N} 
\int\! d^2R \int_0^1 \!\frac{dx}{2x(1\!-\!x)^2} \; \frac{S}{\tilde{S}} \;
\phi(s_i) \; \phi(s_f)  \;,
\label{eq:lf1} 
\end{eqnarray} 
where the arguments, $s_i$ and $s_f$, entering the wave functions may be written as:
\begin{eqnarray}
&&s_i=(p\!+\!p_i)^2=
\frac{m_1^2\!+\!p^2_{i\perp}}{1-x}+\frac{m_2^2\!+\!p^2_{\perp}}{x}-P^2_{i\perp}=
\frac{x\,m_1^2+(1\!-\!x)m_2^2
+(\vec{R}\!-\!x\,\vec{P}_{i\perp})^2}{x\,(1-x)}\, , 
\nonumber \\ 
&&s_f=(p\!+\!p_f)^2=
\frac{m_1^2\!+\!p^2_{f\perp}}{1-x}+\frac{m_2^2\!+\!p^2_{\perp}}{x}-P^2_{f\perp}=
\frac{x\,m_1^2+(1\!-\!x)m_2^2+(\vec{R}\!-\!x\vec{P}_{f\perp})^2}{x\,(1-x)} \, .
\nonumber \\ \label{eq:lf2} 
\end{eqnarray} 
The ratios, $\frac{I^0_{\omega}}{\tilde{I}^0_{\omega}}$ 
and $\frac{S}{\tilde{S}}$ have been inserted in the expressions 
of form factors for the case of scalar constituents. They take into account 
that we are dealing here with  spin-1/2  constituents instead of scalar ones. 
They read:
\begin{eqnarray}
&& \frac{I^0_{\omega}}{\tilde{I}^0_{\omega}}=
 \frac{2(1\!-\!x)\Big(\bar{s}\!-\!(m_1\!-\!m_2)^2\Big)+xq^2
}{2(1\!-\!x) \sqrt{s_i\!-\!(m_1\!-\!m_2)^2}\;\sqrt{s_f\!-\!(m_1\!-\!m_2)^2}}\,,
\nonumber \\
&& \frac{S}{\tilde{S}}=\frac{ 2m_1\Big(\bar{s}\!-\!(m_1\!-\!m_2)^2\Big)\!-\!m_2\,q^2   }{
2m_1\sqrt{s_i\!-\!(m_1\!-\!m_2)^2}\;\sqrt{s_f\!-\!(m_1\!-\!m_2)^2}}\, ,
 \label{eq:lf12}
\end{eqnarray}
The above expressions can be cast into the dispersion-relation ones by an
appropriate change of variables, which allows one to reduce the 3-dimensional
integration to a 2-dimensional one \cite{Desplanques:2008fg}. A different
demonstration is given in ref. \cite{Melikhov:2001zv}

\subsection{Normalization}
The normalization is most often associated to a conserved current 
and, in absence of other candidate, it is taken as the charge, $F_1(0)$. 
Starting from eq. (\ref{eq:dp15}), one gets after some algebra 
of which detail is given elsewhere \cite{Desplanques:2009un}:
\begin{eqnarray} 
&&F_1(0)= \frac{1}{N} \int d\bar{s} \;\phi^2(\bar{s}) \,
\frac{\sqrt{\bar{s}^2-2\bar{s}\,(m_2^2\!+\!m_1^2)+(m_2^2\!-\!m_1^2)^2}
}{\bar{s}} \,. \label{eq:dp17}
\end{eqnarray}
It is noticed that the above expression is symmetrical in the exchange 
of the constituent masses, $m_1$ and $m_2$, as expected. 
The symmetry property reflects a similar one for the contribution 
of each constituent, allowing one to sum up their charges 
to get the total charge. 

It is useful to make the relation of this expression with
the expression of the norm in terms 
of the internal variable $\vec{k}$. This can be obtained as follows. 
Using the Bakamjian-Thomas transformation for unequal constituent masses
\cite{Bakamjian:1953kh},  possibly generalized to any form 
\cite{Desplanques:2004sp}\footnote{Factors $e_k$ in eq. (2) of this reference should be replaced 
by $e_{1k,2k}$ depending on the particle and the factor $2e_k$  
in the  next equations (3, 4, 5) should be replaced by $e_{1k}+e_{2k}$.}, 
one can express the $s$ variable entering the wave function $\phi(s)$ as:
\begin{eqnarray} 
s=(p_1+p_2)^2=(e_{1k}+e_{2k})^2\, ,
\end{eqnarray}
where $e_{1k}=\sqrt{m_1^2+k^2}$, $e_{2k}=\sqrt{m_2^2+k^2}$. 
Noticing that the above expression for $s$ implies relations such as:
\begin{eqnarray} 
&&2k=\frac{\sqrt{s^2-2s\,(m_2^2\!+\!m_1^2)+(m_2^2\!-\!m_1^2)^2}
}{\sqrt{s}}\, , \hspace*{0.5cm}
\nonumber \\
&&ds=\frac{2k\,(e_{1k}\!+\!e_{2k})^2}{e_{1k}\,e_{2k}}\, dk \,,
\label{eq:dp19}
\end{eqnarray}
as well as the relation for the wave function 
$\tilde{\phi}(k^2=\vec{k}^2)$ that is useful for our purpose:
\begin{eqnarray} 
\phi(s)=
\tilde{\phi}\Big ( \frac{
s^2-2s\,(m_2^2\!+\!m_1^2)+(m_2^2\!-\!m_1^2)^2}{4s}\Big)
= \tilde{\phi}(k^2=\vec{k}^2)\, ,
\label{eq:dp20}
\end{eqnarray}
the expression of the norm given by eq. (\ref{eq:dp17}) can be cast into 
the following ones in terms of the internal $k$ variable:
\begin{eqnarray} 
&&F_1(0)
=\frac{16\pi^2}{N}\int \frac{d\vec{k}}{(2\pi)^3}\, \tilde{\phi}^2(\vec{k}^2) \;
\frac{e_{1k}\!+\!e_{2k}}{2\,e_{1k}\,e_{2k}}
\nonumber \\
&&\hspace*{1cm}=\frac{8}{N} \int dk\, k^2\, \tilde{\phi}^2(k^2) \;
\frac{e_{1k}\!+\!e_{2k}}{2\,e_{1k}\,e_{2k}}
= \frac{8}{N} \int \,dk\, k^2\;\phi_0^2(k)\,. \label{eq:normif}
\end{eqnarray}
This last expression is a rather straightforward generalization of the norm 
for equal-mass constituents. Using the expression  
$k^2=\vec{k}^2=(s^2-2s\,(m_2^2\!+\!m_1^2)+(m_2^2\!-\!m_1^2)^2)/(4s)$ 
(see eq. (\ref{eq:dp19})),  the expression 
$s=(m_1^2\!+\!k^2_{\perp})/(1-x)+(m_2^2\!+\!k^2_{\perp})/x$ 
(see eq. (\ref{eq:lf2})), and the resulting expressions  
$k^z=\Big(x(m_1^2+k_{\perp}^2)/(1-x)-(1-x)(m_2^2+k_{\perp}^2)/x \Big)
/(2\sqrt{s})$,
$dk^z(e_{1k}\!+\!e_{2k})/(e_{1k}\,e_{2k})=dx/(x(1-x))$, 
one can also cast the above expression for the norm into the
following one:
\begin{equation}
F_1(0)
= \frac{16\pi^2}{(2\pi)^3N}\int  \;
\frac{d^2k_{\perp}\;dx}{2x\,(1\!-\!x)} \;\tilde{\phi}^2(\vec{k}^2)\,.
 \label{eq:normff}
\end{equation}
%
\subsection{Pseudoscalar-meson decay constant}
The expression of the pion decay constant obtained in the case 
of an hyperplane with arbitrary orientation $\xi^{\mu}$
\cite{Desplanques:2009} can  be generalized as follows 
to a pseudoscalar meson composed of quarks with different mass:
\begin{equation}
 f_{P} = \frac{\sqrt{3}}{(2\,\pi)^3}\;\frac{4\pi}{\sqrt{N}}
\int  \frac{d\vec{p}}{e_1\;\xi \!\cdot\! p_2}\;\;
 \frac{m_1\,\xi \!\cdot\!\,p_2+m_2\,\xi \!\cdot\! p_1}{
 \sqrt{s-(m_1\!-\!m_2)^2}} \;\tilde{\phi}(\vec{k}^2) \, . 
\label{fpi7}
\end{equation}
For the instant form, it reads:
\begin{equation}
f^{IF}_{P} 
=\frac{\sqrt{3}}{(2\,\pi)^3}\frac{4\pi}{\sqrt{N}} \int 
\frac{d\vec{p}}{e_1\;e_2}\; 
\frac{m_1\,e_2\!+\!m_2\,e_1}{\sqrt{s-(m_1\!-\!m_2)^2}}\; \tilde{\phi}(\vec{k}^2)\,.
 \label{fpi2}
\end{equation}
This expression can be expressed in terms of the internal variable $\vec{k}$ 
and the total momentum $\vec{P}$, using  eqs. (2, 3) 
of ref. \cite{Desplanques:2004sp} (see also footnote 2). 
Taking into account in particular that the integration 
of the quantity $\vec{k}\cdot \vec{P}$ 
over the orientation of $\vec{k}$ gives $0$,
the expression is found to be equal to: 
\begin{equation}
f^{IF}_{P} 
=\frac{\sqrt{3}}{(2\,\pi)^3}\frac{4\pi}{\sqrt{N}} \int \frac{d\vec{k}}{e_{1k}\,e_{2k}}\; 
\frac{m_1\,e_{2k}\!+\!m_2\,e_{1k}}{
\sqrt{(e_{1k}\!+\!e_{2k})^2-(m_1\!-\!m_2)^2}}\;  \tilde{\phi}(\vec{k}^2).
 \label{fpi4}
\end{equation}
The expression is independent of the momentum of the pseudoscalar meson, $\vec{P}$, 
and therefore  is Lorentz invariant. The expression in the ``point form" can be
obtained from eq. (\ref{fpi7}) by using the relevant definition of $\xi^{\mu}$, 
$\xi^{\mu}=\frac{P^{\mu}}{M}$:
\begin{equation}
 f^{``P.F."}_{P} = \frac{\sqrt{3}}{(2\,\pi)^3}\;\frac{4\pi}{\sqrt{N}}
\int  \frac{d\vec{p}}{e_1\;P \!\cdot\! p_2}\;\;
 \frac{m_1\,P \!\cdot\!\,p_2+m_2\,P \!\cdot\! p_1}{
 \sqrt{s-(m_1\!-\!m_2)^2}} \;\tilde{\phi}(\vec{k}^2) \, . 
\label{fpi77}
\end{equation}
This expression is explicitly Lorentz invariant, in contrast 
to the instant-form one where this property is implicit. 
Similarly to this approach however, the ``P.F." expression of the pion 
decay constant can be expressed in terms of the internal variable $\vec{k}$. 
The result so obtained is given by:
\begin{equation}
f^{``P.F."}_{P} 
=\frac{\sqrt{3}}{(2\,\pi)^3}\frac{4\pi}{\sqrt{N}} \int \frac{d\vec{k}}{e_{1k}\,e_{2k}}\; 
\frac{m_1\,e_{2k}\!+\!m_2\,e_{1k}}{
\sqrt{(e_{1k}\!+\!e_{2k})^2-(m_1\!-\!m_2)^2}}\;  \tilde{\phi}(\vec{k}^2).
 \label{fpi44}
\end{equation}
The expression in the front-form case can be found in the literature 
\cite{Melikhov:2001zv} but can also be obtained from eq. (\ref{fpi7}) with a
standard change of variable. It reads:
\begin{equation}
f^{FF}_{P}
=\frac{\sqrt{3}}{(2\,\pi)^3} \frac{4\pi}{\sqrt{N}}\int  \;
\frac{d^2k_{\perp}\;dx}{x\,(1\!-\!x)} \;
\frac{m_1\,x\!+\!m_2\,(1\!-\!x)}{\sqrt{s-(m_1\!-\!m_2)^2}}\;\tilde{\phi}(\vec{k}^2),
 \label{fpi5}
\end{equation}
where the argument of the wave function, $\vec{k}^2$, has been given previously
(see eqs. (\ref{eq:dp19}, \ref{eq:dp20}) and text after eq. (\ref{eq:normif})). 
Not surprisingly, 
the above expression can be recovered from the instant-form one, 
eq. (\ref{fpi2}), in the limit $\vec{P} \rightarrow \infty$.
One has therefore:
\begin{equation}
f_{P}=f^{IF}_{P}=f^{``P.F."}_{P} =f^{FF}_{P}.
\end{equation}
An expression of $f_{P}$ different from the above one, but nevertheless
equivalent, has been given in the
literature \cite{Melikhov:2001zv}\footnote{We are grateful to the author for
confirming that the absence of a factor $\sqrt{x(1-x)}$ in the denominator of
his expression for $f_P$ (eq. (2.65) of the arXiv reference) is a misprint, 
without consequence for numerical results presented in his paper.}. 
With our notations, it reads:
\begin{eqnarray}
f_{P} 
&=&\frac{\sqrt{3}\;(m_1\!+\!m_2)}{(2\,\pi)^3} \frac{4\pi}{\sqrt{N}}
\int \frac{d\vec{k}}{e_{1k}\,e_{2k}}\; 
\frac{\;\sqrt{(e_{1k}\!+\!e_{2k})^2-(m_1\!-\!m_2)^2}
}{2\,(e_{1k}\!+\!e_{2k})}\;  \tilde{\phi}(\vec{k}^2)
\nonumber \\
&=&\frac{\sqrt{3}\;(m_1\!+\!m_2)}{(2\,\pi)^3}\frac{4\pi}{\sqrt{N}}
 \int  \;\frac{d^2k_{\perp}\;dx}{x\,(1\!-\!x)} \;
\frac{\sqrt{s-(m_1\!-\!m_2)^2}}{2s}\;\tilde{\phi}(\vec{k}^2)\, .
\label{fpi6}
\end{eqnarray}
The equivalence of the two expressions can be checked  
by making the  change of variable already mentioned above, 
$k^z=\Big(x(m_1^2\!+\!k_{\perp}^2)/(1\!-\!x)-(1\!-\!x)(m_2^2\!+\!k_{\perp}^2)/x \Big)
/(2\sqrt{s})$.  

Apart from its interest in itself, the decay constant of pseudoscalar-mesons is 
relevant for the description of their form factors. In the pion case, a relation
to the squared charge radius has been proposed in the chiral
symmetry limit \cite{Tarrach:1979,Gerasimov:1979,Bernard:1988,Frederico:1992}:
\begin{eqnarray}
r_{\pi}^2=\frac{3}{4\pi^2 f^2_{\pi}}\, .
\label{eq:rch2}
\end{eqnarray}
The relation, which does not explicitly refer to the pion description, 
holds in the point-like  limit. A second relation concerns the asymptotic
behavior of the pion form factor \cite{Farrar:1979,Lepage:1979zb}:
\begin{eqnarray}
F_1(Q^2)_{Q^2 \rightarrow \infty}= 16 \pi f^2_{\pi} \,\frac{\alpha_s }{ Q^2}\, .
\label{eq:asym}
\end{eqnarray}
Corrections are expected from a non-perturbative calculation \cite{Efrimov:1980}, 
which could also be found in a RQM approach \cite{Desplanques:2009}.

\section{Quantitative effect of constraints for form factors}
We consider in this section the quantitative effect of constraints related to
covariant space-time translations for the charge form factor of
pseudoscalar mesons. This is done for both the pion and the kaon mesons. 
For our purpose, it is sufficient to consider the simplest description 
of the mesons. It is not totally arbitrary however and we assume that
it is given by a Gaussian wave function with a parameter as obtained from a
standard confining potential \cite{Basdevant:1986}:
\begin{eqnarray}
\phi_0(\vec{k})\propto exp(- \vec{k}^2 (1+m_{<}/m_{>})/(2\sigma_{st}))
\label{eq:wf}
\end{eqnarray}
where $\sigma_{st}$ represents the string tension
($\sigma_{st}=0.2$ GeV$^2$$\simeq$1 GeV/fm), while $m_{<}$ and $m_{>}$ 
respectively represent the smallest and largest constituent-quark masses.
The  quark masses are determined by
requiring that they allow one to reproduce approximately the pion 
or kaon decay constants (respectively 0.0924 GeV and 0.113 GeV 
in our conventions for the definition of these constants). 
As mentioned at the end of the previous section, 
there is in the pion case some approximate relation 
of this constant to the charge radius (see eq. (\ref{eq:rch2})).
The above relationship may be therefore relevant for describing 
the pion charge form factor at small $Q^2$. The values we use for the quark
masses are $m_{u,d}=0.25$ GeV and $m_s=0.47$ GeV, corresponding 
to pion and kaon decay constants of 0.092 GeV and 0.113 GeV. 
To the extent where the string tension we are using is determined 
by fitting Regge trajectories, results presented in this section 
are essentially parameter free.

A Gaussian wave function has been used in many works  
\cite{Chung:1988mu,Frederico:1992,Cardarelli:1994,Choi:1997iq,Allen:1998hb,Melikhov:2001zv,Krutov:2002nu,Amghar:2003tx,He:2004ba,Coester:2004qu,He:2005} 
but, in most cases, the parameter on which it depends and the quark masses 
were fitted to the measured pion (or kaon) 
form factor. Though we show calculated form factors and measurements 
in the same figure, our intent in this section is not to compare them in detail.
We stress that our main intent here is to examine the role of constraints 
related to space-time translations and determine which approach 
is, ultimately, the most appropriate for a comparison to measurements. 
Nevertheless, a comparison at this point may be useful to  tell us how good 
are estimates based on a wave function whose parameters are obtained 
from considerations different from the measured form factors. It can determine
whether there is a large space for improvement from considering a better wave
function to be considered in the next section. 

Charge form factors of the charged pions and kaons are calculated 
for both a small range of $Q^2$ ($Q^2<0.2$ GeV$^2$), 
which is mainly sensitive to the charge radius, and an intermediate range 
which could be of relevance for present and future measurements,
($0<Q^2<10$ GeV$^2$). In this case, the form factors are multiplied by a
factor $Q^2$, as it is expected that this product should tend asymptotically 
to a constant (up to log terms). The contribution that could account for 
this asymptotic behavior within a RQM framework \cite{Desplanques:2009} 
is however ignored in the present section but will be considered in the next one.

Different RQM approaches are considered but, in all cases, calculations are
performed in the Breit frame. They involve the standard front-form
one with $q^+=0$ (denoted F.F. (perp.), the instant-form one (denoted I.F.), 
a front-form one with the momentum transfer oriented along the front
direction, $\vec{n}$ (denoted F.F. (parallel)) and a point-form one (denoted
``P.F."). We notice that this point form differs from  the Dirac one 
based on a hyperboloid surface \cite{Sokolov:1985jv}. 
As mentioned  by Bakamjian 
\cite{Bakamjian:1961}, it is rather some kind of instant-form approach
with the symmetry properties of Dirac's point-form. Contrary to some
previous papers \cite{Desplanques:2004sp,Desplanques:2008fg}, 
we do not present results for an approach inspired from 
Dirac's point-form \cite{Desplanques:2004rd}. 
These ones, which involve front-form ones with a summation over all
directions (making their calculation rather lengthy), drop between the two
cases shown here for a perpendicular and a parallel configuration of the
momentum transfer, $\vec{q}$, and the front direction, $\vec{n}$. We could
also add that the two front-form calculations presented here respectively 
coincide with the instant-form ones in an infinite momentum frame 
with a momentum transfer perpendicular and parallel to this infinite momentum. 

\begin{figure}[htb]
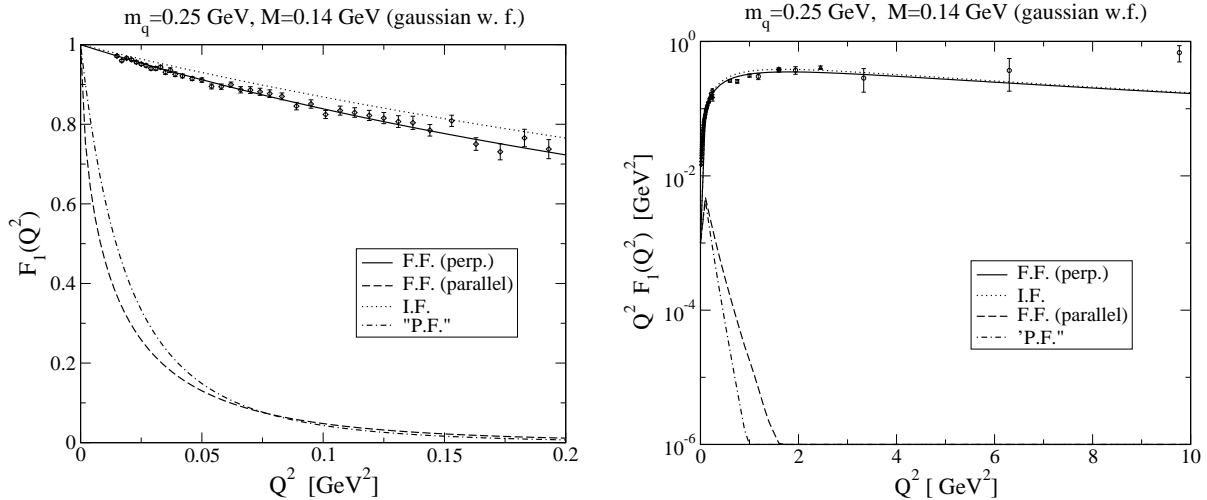
 
\epsfig{file=pions.eps, width = 7.6cm}  \hspace*{0.3cm}
\epsfig{file=pionss.eps, width = 7.7cm} 
\caption{Pion charge form factor, $F_1(Q^2)$,  at small and high $Q^2$ 
(left and right panels respectively). In the latter case, the form factor is
multiplied by $Q^2$ and represented on a logarithmic scale 
to emphasize the asymptotic behavior. The different curves represent
form factors without accounting for constraints from covariant space-time
translations except for the F.F. (perp.) one (continuous line) that satisfies
these properties. When constraints are accounted for, all curves coincide with
the F.F. (perp.) one. All form factors are calculated in the Breit frame. 
\label{PI1}}
\end{figure} 

Expressions of form factors that are used here have been given in a paper
summarizing the technical aspects \cite{Desplanques:2009un}. 
It is reminded that these expressions  ensure that form factors are boost 
and rotation independent and lead to the same results after accounting 
for constraints related to the covariant transformations of currents 
under space-time translations. As these results were given 
for quite general cases, we give in the appendix \ref{app:B} 
the explicit expressions for the Breit-frame case considered 
in the present work. In this respect,  we note that expressions of form factors
used here differ from the ones used in ref. \cite{Desplanques:2009}, 
except for the front-form case with $q^+=0$. This work, mainly devoted 
to the asymptotic behavior of the pion charge form factor, was not concerned 
with accounting for the constraints related to covariant space-time translations. 
Though the numerical effect is a relatively minor one, the choice of the current
adopted there differs from the one used here at $Q^2\neq0$. 
This could allow one to get results close to each other for different approaches 
after taking into account the constraints discussed in this work 
but would prevent one to get identical results for the different approaches 
as here. 
Let's add that the form of matrix elements of the single-particle currents 
used here was suggested by an analysis of expressions obtained 
with the simplest Feynman triangle diagram  \cite{Desplanques:2008fg} 
(and taking into account that we are dealing in this work with fermions 
instead of bosons).

\begin{figure}[htb]
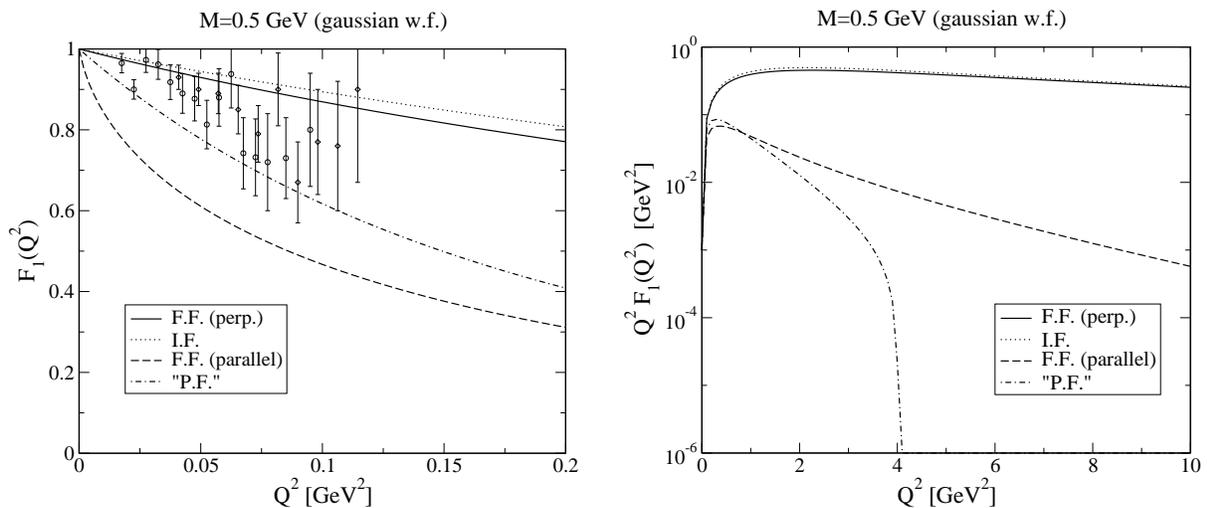
 
\epsfig{file=kaons.eps, width = 7.6cm}  \hspace*{0.3cm}
\epsfig{file=kaonss.eps, width = 7.7cm} 
\caption{Same as in fig. \ref{PI1}, but for the kaon meson. \label{KA1}}
\end{figure}

Results for the pion and kaon charge form factors in absence of the
constraints related to space-time translations are shown in figs. 
\ref{PI1} and \ref{KA1} respectively, together with the corresponding
measurements (refs. \cite{Bebek:1978pe,Amendolia:1986wj,Volmer:2000ek,Horn:2006,Tadevosyan:2007,Huber:2007} 
for the pion and refs. \cite{Daily:1980,Amendolia:1986} for the kaon).
Examination of the figures shows that front-form form factors in the perpendicular
configuration ($q^+=0$) and  the instant-form ones are close to each other.
They strongly differ from the other ones (front-form ones in the parallel
configuration and point-form ones), especially for the pion where discrepancies
can reach many orders of magnitude in the high-$Q^2$ range. This result can be
related to the departure of the momentum transferred to the constituents,
$(p_i-p_f)^2$, from
the one transferred to the system under consideration, $q^2$, preventing one 
to satisfy eq. (\ref{eq:translat7}) that is expected for Poincar\'e covariant
space-time translations. Looking at eq. (\ref{eq:q2}), 
it is found that the discrepancy in the first case, due to the vanishing of
the factor $q \cdot \xi$,  involves terms that are of the
second order in interaction effects. In the second case, this factor does not 
vanish and first order effects are therefore  involved. The steep slope 
of the pion form factor at low $Q^2$ can be then related to the appearance 
of a factor of the order $4e_k^2/M^2$ in front of the $Q^2$ term 
in the argument of wave functions entering the corresponding
expressions for form factors. As the pion mass is small in comparison of the
sum of the constituent masses, the effect due to this factor is necessarily
large. In comparison, the kaon mass is closer to the sum of the constituent
masses and, as a result, discrepancies are significantly smaller.

Not surprisingly, present results show a strong qualitative similarity 
with those for scalar constituents \cite{Desplanques:2008fg}. The effects
mentioned here essentially involve the wave functions and are relatively
insensitive to the current operator. There is however one effect that the
restriction of the present study to charge form factors does not allow us to
evidence. In comparison with the scalar constituent case, the charge form
factor is systematically suppressed with respect to the scalar form factor at
high $Q^2$. This effect, which involves the current operator, 
is roughly given by a factor $1/Q^2$.

When constraints related to  covariant space-time translations 
are accounted for, form factors as those shown in figs. \ref{PI1} 
and \ref{KA1} tend to get closer to each other \cite{Desplanques:2004sp}. 
In the present case, the choice of the currents in different forms 
also ensures Lorentz invariance \cite{Desplanques:2009un}. 
Thus, accounting for the above constraints makes form
factors shown   in figs. \ref{PI1} and \ref{KA1} identical to the front-form
ones in the perpendicular configuration ($q^+=0$, denoted F.F. (perp.)). 
There is therefore no need to present new figures. It is worth noticing that
accounting for the constraints has removed tremendous discrepancies at both
low and large $Q^2$. In particular, the steep slope of the pion charge form 
factor at low $Q^2$, which could be infinite when the pion mass vanishes,  
has disappeared. In this case, the constraints allow one to get rid of the
paradox where the charge radius becomes infinite while the mass of the system
goes to zero, what is generally obtained by increasing the attraction. 
More generally, the constraints can reduce or even remove ambiguities 
like the one which leads to get very different form factors from 
employing a unique wave function, depending  on the mass of the system, 
while one would expect a unique result from a non-relativistic calculation. 
To some extent, the constraints restore fundamental symmetry properties 
that are missed from using some particular (truncated) approaches, somewhat
similarly to what occurs in other fields of physics.

\begin{figure}[htb]
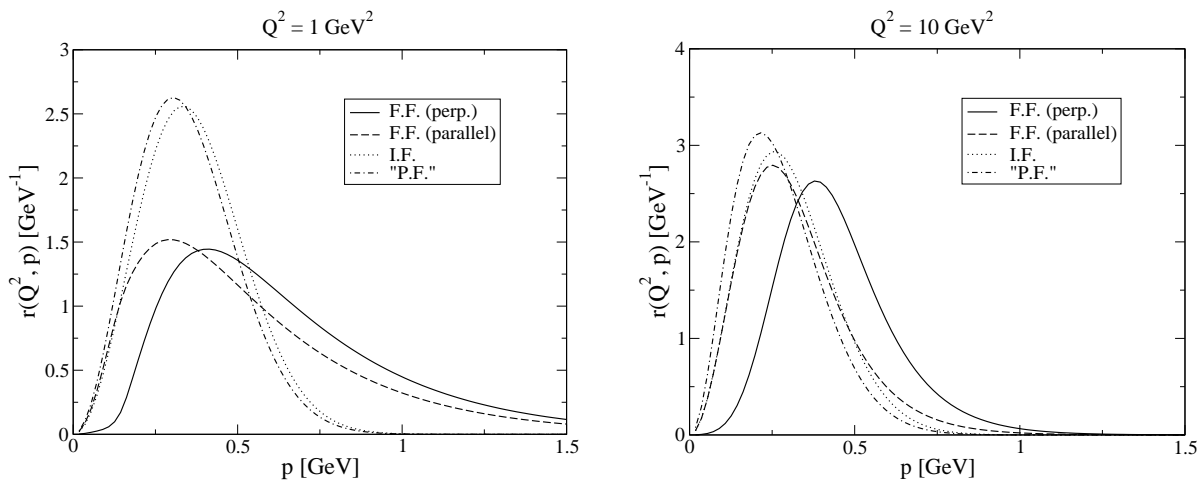
 
\epsfig{file=ffx1.eps, width = 7.6cm}  \hspace*{0.5cm}
\epsfig{file=ffx10.eps, width = 7.45cm} 
\caption{Contribution to the pion charge form factor as a function of the
spectator momentum, $p$. Results  represent the ratio $r(Q^2,p)$ 
of the integrands to the integrated form factor (the same for all cases). 
The integral of these ratios over $p$ is equal to 1.\label{PI2}}
\end{figure} 
It has been shown that accounting for the constraints related to  covariant 
space-time translations could allow one to cast the form factors calculated 
in different forms into a unique expression that is suggested by using a
dispersion-relation approach. While expressions for form factors then involve
the same integrand, we would like to stress that this result is far from being
trivial. It is obtained by using a change of variables that differs from one
form to the other. This is evidenced by looking at the contribution to form
factors corresponding to given values of the momentum of the spectator
constituent, which in the present work, is chosen as the integration variable.
For illustrating this result, we considered the integrand, $f(Q^2,p)$,  
obtained in different forms after integrating on the components 
of the spectator momentum corresponding to a total momentum, 
$p=\sqrt{p^2_{\parallel}+p^2_{\perp}}$. 
The ratios of these integrands to the form factors, which are defined as:
\begin{equation}
r(Q^2,p)= \frac{f(Q^2,p)}{\int \! dp\,  f(Q^2,p)}\, ,
\label{eq:ratio}
\end{equation} 
are shown in fig. \ref{PI2} as a function of the $p$ variable 
for two momentum transfers, $Q^2=1$ and $Q^2=10$ GeV$^2$. 
As the examination of
the two panels shows, results  depend significantly on the approach, as well
as the momentum transfer. It is noticed that results at low  $Q^2$ fall 
in two sets that are indicative of what occurs at the value $Q^2=0$  
(not shown in the figure) where the two front-form results and  
the instant- and point-form separately coincide with each other. The
contribution to the form factor extends to large momenta  in the first case 
while it is rather concentrated at relatively small momenta in the second case. 
At larger  $Q^2$, the contribution to the form factor tends to concentrate to
the lower $p$ range in all cases. It sounds that the results 
for the instant-form and  front-form one with the parallel momentum 
configuration are getting closer to each other and could coincide in the
infinite $Q^2$ limit. As the first approach  in an infinite-momentum frame 
coincides with the second one, such a result may not be so surprising.

In this section, we concentrated on the role of constraints related to covariant
space-time translations. Having shown that accounting for them amounts to obtain
form factors equal to the front-form ones with $q^+=0$ (F.F. (perp.)), we can
consider these last form factors for a first comparison with measurements. While,
there is essentially no free parameter, it is found that there is a rough
agreement in the pion case. 
It is noticeable that the calculated pion charge form factor evidences some kind
of plateau in the range 1--5 GeV$^2$. As mentioned previously, this plateau
has nothing to do with the theoretically expected one as the corresponding
physics has been ignored. It simply results from the fact that the product of the
form factor by $Q^2$ has a maximum in the above range. With this respect, the
departure for the point around 10 GeV$^2$ probably provides hint for missing
physics.
In the kaon case, there could be some trend to
slightly overestimate the experimental data. In view of this first comparison
with measurements, improving the
description of pseudoscalar meson form factors could reveal to be difficult.

\section{Reproducing measurements}
In this section, we intend to make some comparison with measurements, taking
into account that form factors can be identified to those obtained 
in the standard front-form case with $q^+=0$ after the effect of constraints
related to covariant space-time translations has been incorporated. 
The comparison mainly concerns the pion charge form factor, 
which has been measured in a momentum-transfer range larger 
than for the kaon case. We consider successively: further physical ingredients
that should be accounted for, the perturbative calculation of the solution 
of the mass operator, and results so obtained once the QCD coupling $\alpha_s$ 
is given some standard value. The discussion essentially involves 
the value of the string tension, $\sigma_{st}$, and the quark mass, $m_q$. 
In most cases, we impose that the pion-decay constant be reproduced, which
determines the value of the last quantity in terms of $\sigma_{st}$ 
and the physical ingredients under consideration. The pion decay constant 
indeed enters in a chiral-symmetry calculation that provides a large part 
of the squared pion charge radius.

\subsection{Further physical ingredients}
To make a relevant comparison with measurements, 
further physical ingredients should be considered besides the contribution 
to the mass operator of the only confining force we accounted for 
in the previous section for simplicity. We thus expect  some contribution 
due to the one-gluon exchange force. Some was considered in the past
\cite{Cardarelli:1994,Cardarelli:1995dc,Desplanques:2009} but as mentioned in
the introduction, it tends to overestimate the pion charge form factor at high
$Q^2$ with the currently used value of the QCD coupling $\alpha_s \simeq 0.4$.
This suggests that the underlying non-perturbative calculation  needs 
corrections. In the present work, we perform a perturbative calculation
with respect to this one-gluon exchange, which can be therefore considered 
as providing a minimal effect. It is expected that this contribution enhances 
the high-momentum tail of the solution of the mass operator and, consequently, 
produce an increase of the form factor at very high $Q^2$ 
in comparison to the Gaussian solution. The former has a power-law behavior 
($1/Q^4$) \cite{Carbonell:1998,Maris:1998}  while the latter vanishes 
exponentially. 

There is a second contribution involving  one-gluon exchange. As is known, a standard estimate of the pion charge form
factor from a single-particle like current  misses the expected asymptotic behavior 
of this form factor \cite{Carbonell:1998,Maris:1998}. 
To reproduce this behavior ($1/Q^2$), a two-particle one-gluon exchange 
current, which implies the off-mass shell behavior of the one-gluon exchange
interaction, has to be considered. Its expression has been recently determined 
for a RQM calculation of the pion charge form factor \cite{Desplanques:2009}. 
Interestingly, its contribution, 
which involves the low-momentum component of the meson wave function, 
is not depending too much on its description. It provides an increase of the
charge form factor corresponding to a small decrease 
of the squared charge radius ($\Delta r^2_{ch}\simeq -0.07$ fm$^2$) 
and an increase of the product $Q^2\,F_1(Q^2)$ 
(of the order of 0.2 GeV$^2$ at $Q^2=10$ GeV$^2$). 
The first contribution looks small in comparison to the total squared charge
radius but is not so small if it is compared to the part of this quantity that
could be attributed to the matter squared radius (as roughly obtained 
from the difference between the measured one, 0.43 fm$^2$, and the part 
given by a chiral-symmetry calculation in the point-like limit, 
$3/(4\pi^2 f^2_{\pi})=0.34$ fm$^2$. The second contribution is definitively
relevant at the highest values of $Q^2$ where the pion charge form
factor has been measured.

\subsection{Perturbative calculation for the solution  of the mass operator}
To obtain a perturbative solution of the mass operator, we start from eq. (1)
of ref.  \cite{Desplanques:2009}, which was given for a non-perturbative case. 
Due to the non-local character of the interaction, the solution is more easily
calculated in momentum space than in configuration space. The correction to the 
unperturbed Gaussian solution, $\phi_0(\vec{k})$, is given for the free Green's 
function case by: 
\begin{eqnarray}
\Delta \phi_0(\vec{k})=\frac{1}{4e^2_k-M^2}\int \; \frac{d\vec{k}'}{(2\,\pi)^3} \;
\frac{4\;g_{eff}^2\;\frac{4}{3}\;e_k\;e_{k'}
\Big(2- \frac{m^2_q\,(e^2_k+e^2_{k'})}{2\,e^2_k\;e^2_{k'}}\Big)}{
\sqrt{e_k}\;\;(\vec{k}-\vec{k}')^2\;\;\sqrt{e_{k'}}}\;\phi_0(\vec{k}')\,,
\end{eqnarray}
where $g_{eff}^2$ is related to the strong coupling  $\alpha_s$ 
by the relation $g_{eff}^2/(4\pi)=\alpha_s$. 
The model implicitly used for describing the confinement interaction 
in the previous section (harmonic-oscillator-type interaction)  allows one 
to do better however by incorporating its effect in the Green's function. 
The solution then reads:
\begin{eqnarray}
&&\Delta \phi_0(\vec{k})=\sum_n \frac{\bar{\phi}_n(\vec{k})}{(6\!+\!8n)\sigma_{st}+4m^2_q\!-\!M^2}
\nonumber \\
&& \hspace*{2cm}\times \int \!\int 
\frac{d\vec{k}"}{(2\,\pi)^3}\;
\frac{d\vec{k}'}{(2\,\pi)^3}\;\bar{\phi}_n(\vec{k}')\;
\frac{4\;g_{eff}^2\;\frac{4}{3}\;e_{k"}\;e_{k'}
\Big(2- \frac{m^2_q\,(e^2_{k"}+e^2_{k'})}{2\,e^2_{k"}\;e^2_{k'}}\Big)}{
\sqrt{e_{k"}}\;\;(\vec{k"}-\vec{k}')^2\;\;\sqrt{e_{k'}}}\;\phi_0(\vec{k}')\,,\;\;\;\;
\end{eqnarray}
where $\bar{\phi}_n(\vec{k})$ describes the radial excitations given by a
harmonic oscillator model consistently with the Gaussian solution employed 
for the ground state 
(normalization $ \int \frac{d\vec{k}}{(2\pi)^3}  \;\bar{\phi}_n^2(\vec{k})=1$). 
The modification of the Green's function mainly
affects the low-$k$ behavior of the correction to the solution of the mass
operator, which it tends to reduce. In practice, we add to the correction given
in the free Green's function case the correction implied by the more complete 
Green's function with $n \leq 500$.

\subsection{Results and discussion}
\begin{figure}[htb]
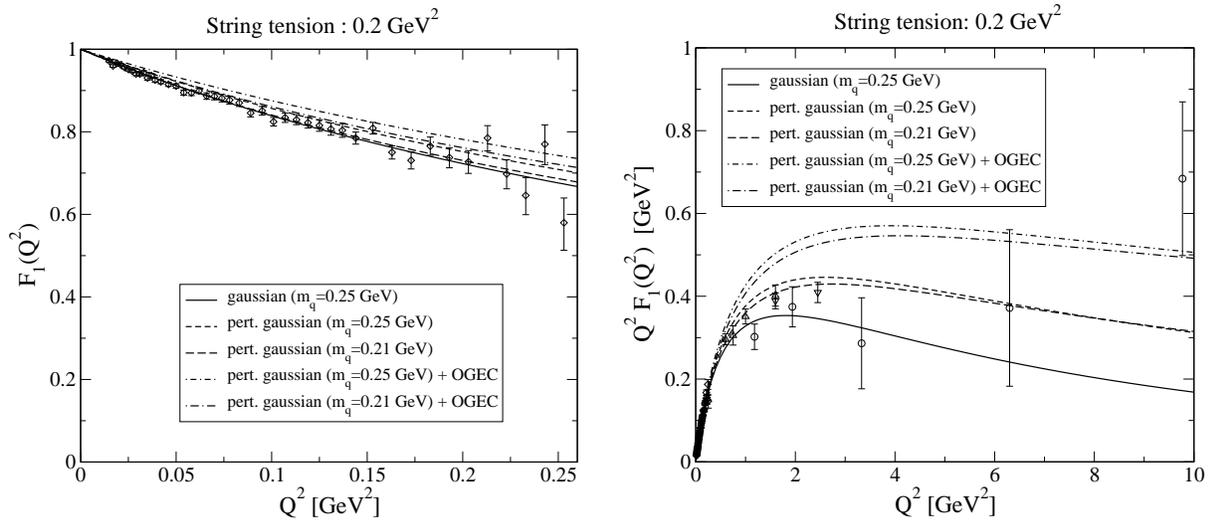
 
\epsfig{file=pionds.eps, width = 7.6cm}  \hspace*{0.3cm}
\epsfig{file=piondss.eps, width = 7.75cm} 
\caption{Pion charge form factor, $F_1(Q^2)$,  at small and high $Q^2$ 
(left and right panels respectively). In the latter case, the form factor is
multiplied by $Q^2$ and represented on a linear scale 
to emphasize a possible $1/Q^2$ asymptotic behavior. 
All curves correspond to the string tension $\sigma_{st}=0.2$ GeV$^2$. 
The different curves represent
form factors corresponding to a Gaussian solution for the mass operator, 
the solution including the perturbative one-gluon exchange in the interaction, 
this last solution with including the contribution of the two-particle current 
and, finally, these last results with a quark mass fitted to reproduce the pion
decay constant $m_q=0.21$ GeV instead of $m_q=0.25$ GeV in the previous results.
\label{compar20}}
\end{figure} 
We first present results by adding to those for the pion charge form factor 
given in the previous section the contribution involving the one-gluon
exchange contribution from a perturbative calculation as described above 
and the contribution due to two-particle currents as described 
in ref. \cite{Desplanques:2009} for the standard front-form case. 
The intent is to provide us with a qualitative illustration of the effect 
of these contributions. This is done in fig. \ref{compar20} 
where we show the form factors obtained with a Gaussian solution 
of the mass operator (previous section, continuous line), 
and with adding successively the effect of a one-gluon exchange 
in the quark-quark interaction (short-dash line) 
and of a two-particle current (short-dash dotted line, denoted OGEC). 
As these last two calculations correspond to a different value 
for the pion decay constant, $f_{\pi}$, 
we also show the  results with  the right pion decay constant 
(long-dash and long-dash dotted lines), which supposes to modify the
quark mass from $m_q=0.25$ GeV to $m_q=0.21$ GeV. 
The results without the two-particle current could be compared to the results
involving the confinement interaction only. 
As in the previous section, results for the form factor extending 
up to $Q^2=10$ GeV$^2$ are multiplied by the factor $Q^2$, 
in relation with the expectation of a plateau in the asymptotic limit,  but, 
as the logarithmic scale does not justify any more here, we adopt a linear
scale.

Corrections due to the one-gluon exchange in the interaction 
provide {\it a priori} a sizeable effect. However, at low $Q^2$,
the correction is almost proportional to the form factor without 
the one-gluon exchange and, as examination of the corresponding panel shows, 
not much effect is seen after renormalizing the form factor at $Q^2=0$ to its
conventional value 1 (more than a factor 2). 
The slight discrepancy that can be observed between the continuous 
and short-dash lines can be traced back to the fact that including 
the one-gluon exchange in the interaction modifies the value of $f_{\pi}$ 
from 0.092 to 0.1015 GeV.
As there is some indication that part of the squared charge radius is
proportional to the inverse of the square of the pion decay constant, 
see the relation given by eq. (\ref{eq:rch2}) and ref. \cite{Frederico:1992} 
for some probing of this relation, we examined the consequences 
due to a change of this value. 
As the comparison of the continuous and long-dash lines indicates, 
it is found that the discrepancy can be largely removed by using 
in the calculation that includes corrections due to the one-gluon
exchange in the interaction a quark mass consistent with the value  
$f_{\pi}=$ 0.092 GeV, ($m_q=$ 0.21 GeV instead of 0.25 GeV). 
The correction due to the one-gluon exchange current, 
being proportional to $Q^2$, does not represent much in the low-$Q^2$ range. 
Its size nevertheless compares to the discrepancies between different calculations 
of the contributions to form factor due to the single-particle current.

Significant effects begin to show up at higher $Q^2$ as examination of the
corresponding panel in fig. \ref{compar20} indicates. The correction due
to the one-gluon exchange in the interaction (discrepancy 
between the continuous and short- or long-dash lines) tends to show a plateau 
in the range (5--10) GeV$^2$ while the form factor without this contribution
begins to decrease. Actually, this plateau is misleading and it rather represents
a maximum as calculations at much higher $Q^2$ shows. The appearance of a plateau
in the product of the form factor by  $Q^2$, as expected from QCD, 
is produced by the one-gluon exchange current contribution (OGEC), which, 
due to its proportionality to $Q^2$ at small values of this quantity, 
is still increasing in the range considered in the figure 
(the value at $Q^2=10$ GeV$^2$ is off the asymptotic value by about 20\%).
Considering the total form factor, it is found to significantly overshoot
measurements as in the low-$Q^2$ domain but, contrary to this case, the
discrepancy cannot be alleviated by modifying the quark mass so that to
reproduce the pion decay constant, as shown in the figure. 
With this respect, it is noticed that the slight discrepancy 
between the two calculations of the contribution of the two-particle current 
around $Q^2=10$ GeV$^2$ has some relationship with the discrepancy 
between the corresponding pion decay constants, as expected from the expression
of the asymptotic form factor given by eq. (\ref{eq:asym}). 
Thus, the consideration of the various contributions due to one-gluon exchange 
on top of the confinement interaction improves to some extent the description 
of the pion form factor at high $Q^2$ but probably provides a too large effect.
\begin{figure}[htb]
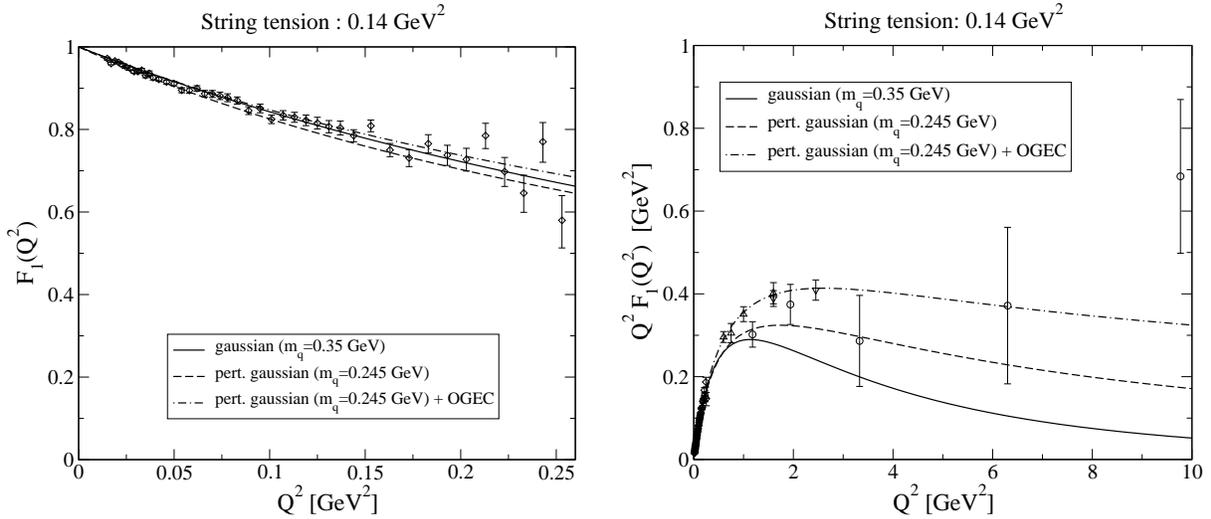
 
\epsfig{file=pionfs.eps, width = 7.6cm}  \hspace*{0.3cm}
\epsfig{file=pionfss.eps, width = 7.75cm} 
\caption{Pion charge form factor, $F_1(Q^2)$,  at small and high $Q^2$ 
(left and right panels respectively). In the latter case, the form factor is
multiplied by $Q^2$ and represented on a linear scale 
to emphasize a possible $1/Q^2$ asymptotic behavior. 
All curves correspond to the string tension $\sigma_{st}=0.14$ GeV$^2$ and assume
the same value of the pion decay constant. 
The different curves represent form factors corresponding 
to a Gaussian solution for the mass operator ($m_q=0.35$ GeV), 
the solution incorporating the perturbative one-gluon exchange in the interaction 
($m_q=0.245$ GeV), and this last solution with including the contribution 
of the two-particle current (OGEC).
\label{compar14}}
\end{figure} 

The second set of results is intended to remedy the above high-$Q^2$ discrepancy
and implies a change of the string tension $\sigma_{st}$ so that to approximately reproduce 
measurements. In this case, the quark mass is always fitted to
reproduce the pion decay constant, $f_{\pi}$, and corresponding results 
are presented in fig. \ref{compar14}. We considered many values of the string
tension and decrease it down to 0.14 GeV$^2$, so that to approximately reproduce
the most accurate measurements in the range $Q^2=$1--2 GeV$^2$. As the
corresponding panel shows, the one-gluon exchange contributions in both the
interaction and in the current begin to be relevant in this range before
becoming essential at higher values of $Q^2$. Due to large error bars, 
it is difficult to say whether the present calculation disagrees with the
measurement at the point close to $Q^2=10$ GeV$^2$. At low $Q^2$ 
(left panel of fig. \ref{compar14}), it is found that the three calculations 
are close to  measurements at first sight. A closer examination nevertheless
shows that the calculation with a Gaussian wave function or the one
incorporating contributions from one-gluon exchange in both the interaction and
in the current are doing better as far as the squared charge radius is concerned.
In units of fm$^2$, the numbers are 0.41 and 0.42 respectively instead of 0.49 
in the third case while the  usually value quoted from measurements is 0.43. 
The comparison with the theoretical numbers for the string tension 0.20 GeV$^2$, 
0.45, 0.39 and 0.46 respectively, shows a significant effect from one-gluon
exchange contributions but, also, a significant dependence on the strength 
of the confinement. This one implies a different balance of contributions 
due to the one-gluon exchange in the interaction and in the current.
\begin{figure}[htb]
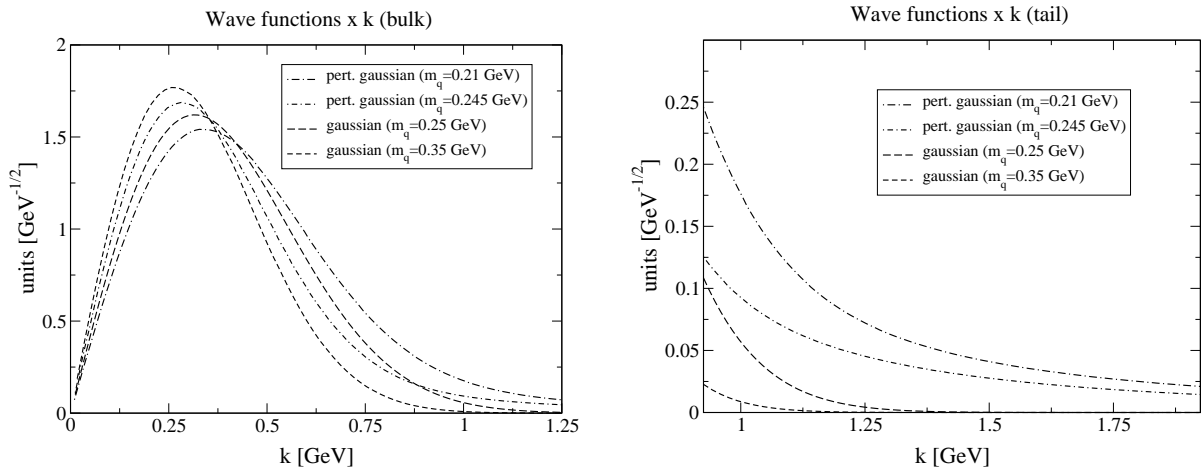
 
\epsfig{file=wfs1.eps, width = 7.6cm}  \hspace*{0.4cm}
\epsfig{file=wfs2.eps, width = 7.6cm} 
\caption{Products of the solutions of the pion mass operator considered 
in this work with the factor $k$, $k\,\phi_0(k)\sqrt{8/N}$ 
(normalization $\int dk \,\Big(k\,\phi_0(k)\sqrt{8/N}\Big)^2=1 $). Results are shown as a function of the internal
$k$ variable  for the bulk contribution ($k \leq 1.25$ GeV, left panel) 
and for the tail contribution ($k \geq 0.925$ GeV, right panel). 
They involve different values of the string tension ($\sigma_{st}=0.2$ GeV$^2$  
and $\sigma_{st}=0.14$ GeV$^2$), without and with incorporating the effect 
of a perturbative one-gluon exchange contribution. In all cases, the quark mass
is fitted to reproduce the pion decay constant. \label{comparwf}}
\end{figure} 

To complement the above results, we also give in fig. \ref{comparwf} 
the different solutions of the mass operator used here: 
Gaussians with $\sigma_{st}=$0.2 GeV$^2$ and 0.14 GeV$^2$, 
and the corresponding perturbative calculations. In all cases, the quark mass 
is fixed to reproduce the pion decay constant, $f_{\pi}=$ 0.092 GeV, 
with the result $ m_q=$0.25, 0.35, 0.21 and 0.245 GeV respectively. 
The left panel shows the bulk contribution, which is roughly the same 
for all cases. Part of the differences can be ascribed to the choice 
of the string tension and to the  one-gluon exchange contribution that slightly 
shifts the wave function to higher values of the internal variable $k$.
The role of the one-gluon exchange contribution is better seen 
on the right panel. In its absence, as can be seen on the figure, 
wave functions drop exponentially. When it is considered, the tail of the wave
function shows a slow decrease given by the power-law behavior $k^{-7/2}$ 
($k^{-5/2}$ for the quantity shown in the figure). This tail is responsible 
for a slower decrease of form factors in comparison to the pure Gaussian case.

Due to a limited experimental information, we did not consider the kaon charge
form factor. Let's first mention that a lower value of the string tension, as
suggested by the pion results, would provide a better account of measurements
in comparison to what was shown in sect. 5. The main question is whether the
one-gluon exchange contribution to the current would cancel most of the effect
as it does for the pion. In absence of a detailed study of this contribution
for the kaon, it is difficult to answer. We only note that a calculation with
an average quark mass, $m=(m_{u,d}+m_s)/2$ leads to an effect 
slightly less than a half of the pion one.

We provided in the text of the present paper a discussion of ingredients
allowing us to make a comparison with other RQM calculations, especially with
respect to the choice of the mass operator,  possible quark form factors 
and non-perturbative effects (introduction and sect. \ref{mass_op}), 
and two-particle currents motivated by the asymptotic behavior 
of the pion charge form factor (this section). 
One could also compare present results with those of field-theory or
lattice calculations. In making such a comparison, one should however take into
account that  calculations to some order in a given framework are not 
necessarily identical to the ones in other frameworks. 
Considering first the contribution that allows one to reproduce the asymptotic
behavior of the pion charge form factor, the present one is found to be 
in good agreement with the field-theory one calculated over a large 
$Q^2$ range in ref. \cite{Bakulev:2004} for instance and with the one  
calculated differently at high $Q^2$ in ref. \cite{Maris:1998} 
(the comparison is not possible at small and moderate $Q^2$ in this case). 
It qualitatively agrees with the one obtained in lattice calculations by
comparing unquenched and quenched calculations \cite{Alexandrou:2006}.
In either case, this contribution is not essential to explain the bulk
contribution at small and moderate $Q^2$ (below $3-4$ GeV$^2$). 
In this range, it looks as
if our results agree with the field-theory or lattice ones but we should stress
that our results are calculated up to the first order in the QCD coupling, 
$\alpha_s=0.4$, 
while the other ones are likely to include many orders. We mentioned possible
problems with accounting for the one-gluon exchange non-perturbatively in a 
RQM calculation together with the standard value $\alpha_s=0.4$. 
The comparison of RQM calculations with field-theory or lattice ones 
probably suggests that the
former ones should use an effective coupling with a value smaller than the
standard one. This could account for retardation effects or, perhaps also, 
the running character of the  QCD coupling. A further suggestion would consist 
in introducing quark form factors.

\section{Conclusion}
In this paper we considered the consequences of covariant Poincar\'e 
space-time translations for the description of the form factor 
of pseudoscalar mesons in a RQM framework. This property is generally used 
to factorize out the dependence on the space-time position 
in the current matrix element, allowing one to obtain 
the energy-momentum conservation. For the calculation of form factors, 
one can thus use  the current at $x=0$, which, most often, is taken 
as a single-particle current, independently of the form of relativity 
that is employed. We showed that the energy-momentum conservation  
does not exhaust all properties stemming from the covariant character 
of Poincar\'e space-time translations. Looking at relations 
derived in the vicinity of the point $x=0$ \cite{Lev:1993}, 
and implying the squared momentum transferred to the constituents 
and to the one transferred to the system, we found that they are 
generally inconsistent with the single-particle assumption for the current. 
Only the standard front-form approach with the condition $q^+=0$ 
could fulfill the expected relations. For other forms 
or other kinematic conditions, we worked out a way to account for the further
relations. It amounts to introduce interaction effects in the relation of the
constituents momenta to the momentum transfer while keeping the single-particle
structure of the current. These effects, that are here or there 
depending on the approach, correspond to take into account 
many-particle currents at all orders in the interaction 
as an expansion would show. In field-theory based approaches, 
they would correspond to Z-type diagram ones.
When these effects are introduced, the equivalence of different approaches 
can be restored provided that the matrix element of the single-particle current 
in each form is appropriately chosen. 

Interestingly, present theoretical results for the pseudoscalar-meson 
charge form factor reproduce those expected in a dispersion-relation approach 
\cite{Melikhov:2001zv,Krutov:2002nu} (with some correction for
the last reference). Except for the standard front-form case  (with $q^+=0$),  
such a result could not be anticipated. Wondering about this result, 
we feel that two ingredients play an essential role. 
To a large extent, the effects of constraints considered here tend to restore 
the equality of the squared momenta transferred to the constituents 
and the one transferred to the system. This relation is always fulfilled 
in a field-theory approach and, {\it a fortiori}, in a dispersion-relation
approach which is based on this framework, but it is generally not satisfied 
in the simplest RQM approach. The second ingredient concerns the matrix element 
of the single-particle current to start with. 
In absence of interaction between constituents, the  scattering amplitude 
of the system with the electromagnetic field which underlies implicitly 
present RQM results, is the same as in the dispersion-relation approach. 
Besides this contribution, the matrix element may also contain 
extra contributions. These ones have an interaction character and ensure 
that its value at $Q^2=0$, which is associated to the charge 
and is not concerned by the above constraints, be Lorentz invariant.
The effective matrix elements so obtained could differ from
the ones naively expected. It thus appears that getting reliable form factors,
which should be independent of the form chosen to implement relativity,
implies all aspects of the Poincar\'e group: boosts, rotations 
but also space-time translations. 

The above effects have been illustrated in the case of the pion and kaon mesons,
using a description of these mesons accounting essentially for the confinement
interaction. The effects can be especially large for some calculations 
in the pion case, in relation with the fact that the pion mass is significantly smaller 
than the sum of the relativistic kinetic energies of the constituents 
in comparison with the kaon case. Noting that in absence of these effects,
these calculations would produce a charge radius that would go to infinity 
while the mass of the system vanishes as well as other paradoxes 
\cite{Desplanques:2003nk}, it is probably not overstated to say 
that the above effects remove violations of space-time translations covariance 
properties that these calculations imply.  
To a very large extent, present results agree qualitatively with those obtained 
for scalar constituents \cite{Desplanques:2008fg} but they offer the advantage 
to concern a physical system. In comparison with experiment, this first set of
results provides a good account of measurements with a string tension equal 
to 0.2 GeV$^2$. It represents a reasonable starting point for improvements.

Necessary improvements involve the contributions of the one-gluon exchange 
in both the interaction and in the current. The first one has been estimated 
from a perturbative-type calculation, taking into account that a complete
non-perturbative calculation could lead to some difficulties 
with the QCD coupling $\alpha_s=0.4$. The second one ensures 
that the expected QCD asymptotic behavior for the charge form factor 
is recovered. It is found in the pion case that the two contributions 
tend to increase the form factor, especially at high $Q^2$. 
A quite reasonable account of measurements is obtained by lowering 
the string tension from 0.2 GeV$^2$ to 0.14 GeV$^2$, a value 
that is probably at the lower limit of what is acceptable. Interestingly, however, 
the change in the string tension has provided a better description of
measurements in both the low-$Q^2$ and the intermediate-$Q^2$ domains. 
We could probably improve the description by playing with the different
ingredients, by requiring for instance that the mass operator reproduces 
the mass difference between pseudoscalar and vector mesons 
or between the non-strange and strange mesons. We nevertheless refrain 
to do it as we believe that many questions could be raised 
at this point concerning the theoretical inputs.

From this study, we can take for granted a number of features. 
The contributions of one-gluon exchange in both the interaction 
and in the current become important ingredients of a realistic description 
of the pion charge form factor around and beyond $Q^2=1 $GeV$^2$, 
due to their increasing role with $Q^2$.  
The role of the contribution in the current could extend to the low-$Q^2$ 
domain where it compares to other effects. Remarkably, this contribution was
found to be rather stable and not depending much 
on which calculation we were performing.
In comparing different results, we found it was important that they correspond 
to the same value of the pion decay constant, what was achieved 
by fitting the quark mass. This prevented us to make biased conclusions 
in many circumstances. While the above contributions seem to go in the right
direction, questions however may arise from the limitation of their study. 

As remembered in the introduction, an unrestricted calculation of the
one-gluon exchange contribution to the interaction would produce a too large
form factor. This could be remedied by using a smaller value for the string
tension but this one is already at the lower limit of what is acceptable.
We should add that describing the confining force as a sum of a kinetic energy 
and a linear $r$-dependent term  could have similar effects to some extent 
and would not {\it a priori} alleviate the problem. This, in turn, raises the
question of the precise form of the confinement interaction and its interplay
with the one-gluon exchange contribution to the interaction. 
Let's mention that retardation effects, which have been considered in the scalar
constituent case, could decrease this last contribution 
but it is not clear whether they would be sufficient to remove the above
problem. They could also affect the calculation of the meson mass spectrum 
but there is no indication that this one will be necessarily improved.
As for the one-gluon exchange contribution to the current, 
the question that may be raised concerns higher-order corrections in the QCD
coupling $\alpha_s$, which have been discarded for simplicity 
in the expression we used. The question also concerns possible corrections related to the
confinement interaction, which have no reason to be negligible 
in comparison to the contribution retained here but are completely unknown.
Ultimately, one could invoke quark form factors 
\cite{Cardarelli:1994,Cardarelli:1995dc} though these ones are not quite
consistent at first sight with the expression of the asymptotic behavior 
of the pion charge form factor. An improved model would consist 
in considering the coupling to the photon of quark-antiquark pairs 
generated by one-gluon exchange. It can incorporate the vector-meson 
phenomenology \cite{deMelo:2007} and in particular the $\rho$-vector 
meson dominance one \cite{Gross:2008} that could be important 
for the range $Q^2\leq$ 1 GeV$^2$. 
Part of this last effect is  accounted for in the present work
\cite{Desplanques:2009} but there is another part corresponding to
quark-antiquark correlations which is certainly not. How much space 
is left for such an effect in the present work is not clear but some 
could allow one to correct or to alleviate part of the drawbacks 
mentioned above.

We started this work with the idea that an implementation of relativity 
free of uncertainties rela\-ti\-ve to the choice of a particular form 
should be used for studying the form factor of pseudoscalar mesons  
and, thus, for getting information on the mass operator describing 
these particles. From a phenomenological viewpoint, it
sounds that this task should not be too difficult, partly due to the fact that
the simplest description is already doing well. The main question is whether
this information is consistent with what is theoretically expected. 
This part of the task could be more difficult. It would probably require 
more elaborate descriptions of both the mass operator and the current 
within the RQM framework but, also, more accurate measurements 
of the pion charge form factor for $Q^2\geq 3$ GeV$^2$ to discriminate 
between different schemes. Clarifying the role of the measurement 
at around $Q^2= 10$ GeV$^2$ by making it more accurate would be 
specially useful. It seems difficult to accommodate its present central value 
within descriptions of the form factor which
do relatively well at smaller momentum transfers.
A complementary study would involve the consideration of the pseudoscalar 
meson spectrum. Though it is likely very approximate, the mass operator 
obtained here could be used for predictions of this spectrum. In turn, the
comparison with experiment could lead to improve the mass operator and help to 
discriminate effects that are due to its description from the ones that are due 
to the description of the electromagnetic interaction.

\noindent
{\bf Acknowledgements}\\
This work is partly supported by the National Sciences Foundations of China
under grants No. 10775148, 10975146, 11035006 (Y.B.). The author is also grateful to the
CAS for grant No KJCX3-SYW-N2. B.D. is very grateful to IHEP and to LPSC 
for offering hospitality allowing him to achieve this work.

\appendix
\section{Relation to the pion charge form factor expressed using 
Wigner rotation angles}
\label{app:A}
In this section, we provide details that allow one to make a comparison with
results for the pion form factor given in ref. \cite{Krutov:2002nu}. We first
give expressions for form factors that include both the Dirac current and an
anomalous magnetic moment so that a more complete comparison can be made with
the above work that contains the two terms (with a different parametrization). 
As this work is based on the use of  Wigner rotation angles while ours does not
consider them explicitly,  we also provide expressions for quantities involving 
the sum of these Wigner rotation angles $\omega_1$ and $\omega_2$, that are
relevant for a comparison. They allow one to get closer expressions for form
factors while considerably simplifying the comparison which shows a discrepancy
factor in the integrand. We finally show how our results, which agree with those
obtained by Melikhov \cite{Melikhov:2001zv}, can be checked from the consideration 
of the simplest
Feynman diagram.

For definiteness, we write the quark current as:
\begin{eqnarray} 
<p_i| j^{\mu}| p_f>= \bar{u}(p_i)\Big(
\tilde{F}^q_1 \,\gamma^{\mu} -\frac{\tilde{F}^q_2}{2m}\, \sigma^{\mu \nu} (p_i\!-\!p_f)_{\nu}
\Big) u(p_f)\,,
\label{eq:app1}
\end{eqnarray}
where $ \sigma^{\mu \nu} =\frac{1}{2}
(\gamma^{\mu}\gamma^{\nu}-\gamma^{\nu}\gamma^{\mu})$.
For our main purpose here, it is sufficient to consider one kind of quark.
It would be straightforward to generalize to different quarks 
expressions given below if necessary.
For on mass-shell quarks, the above current can be cast into the Gordon form:
\begin{eqnarray} 
<p_i| j^{\mu}|p_f>= \bar{u}(p_i)\Big(
 \frac{G_E^q}{2m}\,\frac{(p_i\!+\!p_f)^{\mu}}{1+\frac{Q^2}{4m^2}} 
 + \frac{G_M^q}{(2m)^2}\,\frac{\epsilon^{\mu \nu \rho \sigma} 
 \gamma^{\nu} \gamma_5 (p_i\!-\!p_f)_{\rho} (p_i\!+\!p_f)_{\sigma}
 }{1+\frac{Q^2}{4m^2}}
\Big) u(p_f)\,,
\end{eqnarray}
where: 
\begin{eqnarray} 
G^q_E=\tilde{F}^q_1+\frac{q^2}{4m^2}\tilde{F}^q_2, \hspace*{1cm}
G^q_M=\tilde{F}^q_1+\tilde{F}^q_2\,.
\end{eqnarray}
The corresponding dispersion-relation expression for the pion charge form factor is given by:
\begin{eqnarray} 
&&F_1(Q^2) =\frac{1}{N} \int d\bar{s} \;  d(\frac{s_i\!-\!s_f}{Q}) \; 
 \phi(s_i) \; \phi(s_f)\;
 \nonumber \\
&& \hspace*{2cm}\times\frac{4\tilde{F}^q_1\, s_i\,s_f -\tilde{F}^q_2\,D\,Q^2
 }{2D\sqrt{D}\sqrt{s_i\,s_f}} \;\theta(\cdots)
 \nonumber \\
&& \hspace*{1.5cm}=\frac{1}{N} \int d\bar{s} \;  d(\frac{s_i\!-\!s_f}{Q}) \; 
 \phi(s_i) \; \phi(s_f)\;
 \nonumber \\
&& \hspace*{2cm}\times\frac{G^q_E \,(s_i+s_f+Q^2)^2  
+\frac{G^q_M}{m^2}\,\Big(s_i\, s_f\, Q^2-m^2D\,Q^2\Big)
 }{2D\sqrt{D}\sqrt{s_i\,s_f}\,(1+\frac{Q^2}{4m^2})} \;\theta(\cdots)\, .
\label{eq:app4}
\end{eqnarray}
Though our approach does not explicitly refer to the Wigner rotation angles, 
it is convenient to express some of the factors appearing in the above
expression in terms of these quantities so that to facilitate  the comparison 
with the expression given in  ref. \cite{Krutov:2002nu}. 
In this aim, the following  relations are useful: 
\begin{eqnarray} 
&&{\rm tan}(\omega_1+\omega_2)=\frac{\xi}{ m\,(s_i+s_f+Q^2)}\;,
\nonumber \\
&&{\rm cos}(\omega_1+\omega_2)=
\frac{m\,(s_i+s_f+Q^2) }{ \sqrt{s_i\,s_f\,(4\,m^2+Q^2)}  }\;,
\nonumber \\
&&{\rm sin}(\omega_1+\omega_2)=\frac{\xi}{\sqrt{s_i\,s_f\,(4\,m^2+Q^2)} }\;,
\nonumber \\
&&{\rm with}\;\;\xi=\sqrt{s_i\,s_f\,(4\,m^2+Q^2)-m^2\,(s_i+s_f+Q^2)^2  } \;.
\label{2v}
\end{eqnarray}
One thus gets:
\begin{eqnarray}
&&F_1(Q^2) =\frac{1}{N} \int d\bar{s} \;  d(\frac{s_i\!-\!s_f}{Q}) \; 
 \phi(s_i) \; \phi(s_f)\;
 \nonumber \\
&& \hspace*{2cm}\times 
\frac{G^q_E \,(s_i\!+\!s_f\!+\!Q^2)\,{\rm cos}(\omega_1\!+\!\omega_2) \ 
+G^q_M \,\frac{\xi}{m}\,{\rm sin}(\omega_1\!+\!\omega_2)
 }{D\sqrt{D}\,\sqrt{1+\frac{Q^2}{4m^2}}} \;\theta(\cdots)\, .
\label{eq:app5}
\end{eqnarray}
The contribution to the pion charge form factor for the term $\tilde{F}^q_1$ 
alone, considered in the main text, is easily recovered by using the relation: 
\begin{eqnarray} 
{\rm cos}(\omega_1\!+\!\omega_2)
 +\frac{\xi}{m\,(s_i\!+\!s_f\!+\!Q^2)}\; {\rm sin}(\omega_1\!+\!\omega_2)
 =2\,\frac{\sqrt{s_i\,s_f\,(1+\frac{Q^2}{4m^2})} }{ s_i\!+\!s_f\!+\!Q^2 }\,.
\end{eqnarray}
The comparison of the above expression for the form factor, eq. (\ref{eq:app5}), 
with the one given  in  ref. \cite{Krutov:2002nu} shows that this last one 
contains an extra $Q^2$ dependent factor, $s_i\!+\!s_f\!+\!Q^2$ 
(first line of their eq. (88)), that is not supported neither by our approach, 
nor by Melikhov's one \cite{Melikhov:2001zv} (see eqs. (2.50, 2.53) 
of the web version of the paper). 

The appearance of a discrepancy factor similar to the above one was already
observed for the scalar-constituent case  \cite{Desplanques:2008fg}. 
Actually the full factor is given by 
$(s_i\!+\!s_f\!+\!Q^2)/(2\sqrt{s_i\,s_f})$ so that the charge form factor 
at $Q^2=0$ is unchanged. It could be checked by considering 
the simplest Feynman triangle diagram \cite{Frederico:1992}. 
Due to divergences, the consideration 
of this diagram is not so useful in the spin-1/2 constituent case. 
It nevertheless provides some information for converging contributions 
like the one proportional to the factor $\tilde{F}^q_2$ in eq. (\ref{eq:app1}). 
It is thus found that the contribution to the pion charge form factor can be
cast into the form of the term proportional to $\tilde{F}^q_2$ in eq.
(\ref{eq:app4}) by using the function $\phi(s)=\sqrt{s}/(s-M^2)$. 
This result is inconsistent with the appearance of the extra $Q^2$ dependent
factor, $s_i\!+\!s_f\!+\!Q^2$,  as found in  ref. \cite{Krutov:2002nu}.
A similar conclusion could be obtained for some contributions proportional to 
$\tilde{F}^q_1$ in eq. (\ref{eq:app1}) but caution is required in this case in
separating various diverging terms. The appearance of the extra $Q^2$ dependent
factor, $s_i\!+\!s_f\!+\!Q^2$, in  ref. \cite{Krutov:2002nu} originates from a
factor $P_0 P'_0$ (their eqs. (8) and (28)), which has been calculated by the authors 
in the lab frame where $P_0 P'_0=\frac{1}{2}(s_i\!+\!s_f\!+\!Q^2)$. 
Of course, this result is not Lorentz invariant despite its form. 
The formalism developed by the authors implies in principle 
other factors, $N_C,\; N_{CG}$, that depend on $P_0$ ($ N_{CG}\propto P^2_0$, 
their eq. (7);  $N_C \propto \sqrt{N_{CG}}$, their eq. (10)). 
For a reason unknown to us, these factors have been ignored in their final 
expression for the form factors. We only notice that the combined effect 
of these factors, $(N_C N'_C)/ (N_{CG} N'_{CG})$, in their eqs. (34, 38, 39) 
provides an extra factor  $1/(P_0 P'_0)$ that could cancel 
the above undesirable factor. 
\section{Expressions of form factors used in this work}
\label{app:B}
Expressions of form factors used in this work are obtained from general ones
given in ref. \cite{Desplanques:2009un}. These last ones are somewhat involved
and getting explicit expressions to be used for calculating form factors,
including the effect of constraints related to covariant space-time translations,
may not be straightforward. Giving them here may be therefore useful. In
principle, results do not depend on the frame used for calculations after the
effect related to the above constraints is accounted for but, in practice, it is
necessary to choose some. Results presented in this work have been obtained in
the Breit-frame where some simplification occurs. This is done successively for 
the front form with a momentum transfer perpendicular to the front
orientation (F.F. (perp.), the front form with a  momentum transfer parallel 
to the front orientation (F.F. (parallel)), the instant form (I.F.) 
and the point form (``P.F."). Expressions are given as
integrals over the momentum of the spectator constituent. 
In the front-form case, we consider components,  $p_{\parallel}$ 
and $\vec{p}_{\perp}$, 
that are parallel and perpendicular to the front orientation, $\vec{n}$,  
(defined in terms of the 4-vector \cite{Desplanques:2008fg}  $\omega$ as  $\vec{n}=-\vec{\omega}/\omega^0$). 
In the other cases, we consider  components  that are parallel and perpendicular 
to the momentum transfer, $\vec{q}$. The integrand is given as the product of the one for scalar
constituents multiplied by a factor that represents the ratio of currents 
for spin 1/2 and spin 0 constituents. It involves the variables relative to the
spectator constituent, $p_{\parallel}$ and  $\vec{p}_{\perp}$, as well as other
quantities that depend on them, such as  the argument entering  wave
functions, $k^2_{i,f}$  (or $s_{i,f}$), the energy of the system, $E$, and the
energies of the interacting constituents, $e_{i,f}$. In all cases, the relation
of  $k^2_{i,f}$ to $s_{i,f}$, see eq. (\ref{eq:lf1}), may be written as:
\begin{eqnarray}
k^2_{i,f}=\vec{k}_{i,f}^2 =\frac{1}{4}\,
\Big(s_{i,f}-2(m_1^2+m_2^2)+\frac{(m_1^2-m_2^2)^2}{s_{i,f}} \Big)\,.
\end{eqnarray}
The energy of the system in the Breit frame is given by: 
\begin{eqnarray}
E=\sqrt{M^2+\frac{Q^2}{4}}\,.
\end{eqnarray}
The other quantities, which generally depend on the approach, are given below
for each of them.

\noindent
$\bullet$ Front form with a momentum transfer perpendicular to the front
orientation 
\begin{eqnarray}
&&F_1(Q^2)=\frac{1}{(2\pi)^3} \int  d^2p_{\perp}\,dp_{\parallel}\;
\tilde{\phi}(\vec{k_i}^2)\, \tilde{\phi}(\vec{k_f}^2)\;
\frac{E \;\;\theta(E\!-\!e_p\!-\!p_{\parallel})}{2e_p(E\!-\!e_p\!-\!p_{\parallel})}
\nonumber \\
&& \hspace*{3cm}\times
\frac{ \frac{s_i\!+\!s_f}{2}\!-\!(m_1\!-\!m_2)^2 
-\frac{Q^2\,(e_p+p_{\parallel})}{2(E-e_p-p_{\parallel})}
 }{ \sqrt{s_i\!-\!(m_1\!-\!m_2)^2} \sqrt{s_f\!-\!(m_1\!-\!m_2)^2}} \, ,
\label{eq:ffperp}
\end{eqnarray}
where
\begin{eqnarray}
s_{i,f}=\frac{2e_pE \pm \vec{p}_{\perp}\cdot \vec{q} 
-\frac {e_p+ p_{\parallel}}{E}( E^2\!-\!\frac{Q^2}{4})+m_1^2-m_2^2
}{ 1-\frac{e_p+p_{\parallel}}{E}}\,.
\end{eqnarray}
Using the quantity $x=\frac{e_p+p_{\parallel}}{E}$, it is noticed that the
above expression for the form factor can be cast into the one given by eq.
(\ref{eq:lf1}) while $s_{i,f}$ takes the form given by eq. (\ref{eq:lf2}).

\noindent
$\bullet$ Front form with a momentum transfer parallel to the front orientation 
\begin{eqnarray}
&&F_1(Q^2)=\frac{1}{(2\pi)^3} \int  d^2p_{\perp}\,dp_{\parallel}\;
\tilde{\phi}(\vec{k_i}^2)\, \tilde{\phi}(\vec{k_f}^2)\;
\frac{E\;\;\theta(E\!-\!\frac{\alpha Q}{2}\!-\!e_p\!-\!p_{\parallel})
}{2e_p(E\!-\!e_p\!-\!p_{\parallel})}
\nonumber \\
&& \hspace*{3cm}\times 
\frac { \frac{s_i\!+\!s_f}{2}\!-\!(m_1\!-\!m_2)^2 
-\frac {\!-\!\alpha Q \frac{s_i\!-\!s_f}{2}
-(e_p\!+\!p_{\parallel})(p_i\!-\!p_f)^2}{2(E\!-\!e_p\!-\!p_{\parallel})}
}{ \sqrt{s_i\!-\!(m_1\!-\!m_2)^2} \sqrt{s_f\!-\!(m_1\!-\!m_2)^2}} \, ,
\label{eq:ffpara}
\end{eqnarray}
where
\begin{eqnarray}
&&s_{i,f}=
\frac{(E\mp\frac{\alpha  Q}{2})\Big(2e_pE\pm \alpha  p_{\parallel}Q+m_1^2-m_2^2\Big)
-(e_p\!+\!p_{\parallel})(E^2\!-\!\frac{\alpha^2  Q^2}{4})}{E\mp\frac{\alpha 
Q}{2}\!-\!e_p\!-\!p_{\parallel}}\,,
\nonumber \\
&&\alpha=\frac{E-e_p-p_{\parallel}}{
\sqrt{e^2_p-p^2_{\parallel}+\frac{Q^2}{4}+m_1^2-m_2^2}}\,,
\nonumber \\
&&-(p_i\!-\!p_f)^2=\frac{\alpha^2 Q^2\,(m_1^2 + p^2_{\perp}) }{ 
(E\!-\!\frac{\alpha Q}{2}\!-\!e_p\!-\!p_{\parallel} )
(E\!+\!\frac{\alpha Q}{2}\!-\!e_p\!-\!p_{\parallel}) }\, .
\end{eqnarray}
%

\noindent
$\bullet$ Instant form 
\begin{eqnarray}
&&F_1(Q^2)=\frac{1}{(2\pi)^3} \int  d^2p_{\perp}\,dp_{\parallel}\;
\tilde{\phi}(\vec{k_i}^2) \,\tilde{\phi}(\vec{k_f}^2)\;
\frac{e_i+e_f+2e_p}{2e_p\,(e_i+e_f)}
\nonumber \\
&& \hspace*{3cm}\times 
\frac{ \frac{s_i\!+\!s_f}{2}\!-\!(m_1\!-\!m_2)^2  
-\frac{(e_i\!-\!e_f)\frac{s_i\!-\!s_f}{2} -e_p\,(p_i\!-\!p_f)^2}{e_i\!+\!e_f}
 }{ \sqrt{s_i\!-\!(m_1\!-\!m_2)^2} \sqrt{s_f\!-\!(m_1\!-\!m_2)^2}} \, ,
\label{eq:if}
\end{eqnarray}
where
\begin{eqnarray}
&&s_{i,f}=(e_{i,f}+e_p)^2 -\frac{\alpha^2 Q^2}{4}\,,
\nonumber \\
&&e_{i,f}=\sqrt{m_1^2+p^2_{\perp}+(p_{\parallel}\mp\frac{\alpha Q}{2})^2}\,,
\nonumber \\
&&\alpha=\sqrt{1+\frac{p^2_{\parallel}}{m^2_1+p^2_{\perp}+\frac{Q^2}{4}}}\,,
\nonumber \\
&&-(p_i\!-\!p_f)^2=\alpha^2 Q^2-(e_i\!-\!e_f)^2\,.
\end{eqnarray}
%
\noindent
$\bullet$ ``Point form" \\
The expression for form factors in the  ``point-form" approach is perhaps more 
naturally written in terms of the velocity of the initial or final system 
in the Breit frame, $\mp \vec{v}=\vec{Q}/\sqrt{4M^2+Q^2}$ with our conventions:
\begin{eqnarray}
&&F_1(Q^2)=\frac{1}{(2\pi)^3} \int  d^2p_{\perp}\,dp_{\parallel}\;
\tilde{\phi}(\vec{k_i}^2) \,\tilde{\phi}(\vec{k_f}^2)\;
\frac{1}{e_p\,\Big(1\!-\!\frac{m_2^2-m_1^2}{2s_i}\!-\!\frac{m_2^2-m_1^2}{2s_f}\Big)}
\nonumber \\
&& \hspace*{3cm}\times 
\frac { \frac{s_i\!+\!s_f}{2}\!-\!(m_1\!-\!m_2)^2 
-\frac{(e_i\!-\!e_f)\frac{s_i\!-\!s_f}{2} -e_p\,(p_i\!-\!p_f)^2}{e_i\!+\!e_f}
\Big)
 }{ \sqrt{s_i\!-\!(m_1\!-\!m_2)^2} \sqrt{s_f\!-\!(m_1\!-\!m_2)^2}} \, ,
\label{eq:pf}
\end{eqnarray}
where
\begin{eqnarray}
&&s_{i,f}=(e_{i,f}+e_p)^2(1-\beta^2 v^2)\,,
\nonumber \\
&&e_{i,f}=\frac{e_{\mp}\mp\beta v\,(p_{\parallel}\mp\beta v\, e_p)  
}{1-\beta^2 v^2}\,,
\nonumber \\
&&e_{\pm}=\sqrt{(m_1^2-m_2^2)(1-\beta^2 v^2)
+(e_p \pm \beta v\,p_{\parallel} )^2}\, ,
\nonumber \\
&&\beta v=
\frac{\frac{Q}{2}}{\sqrt{m_2^2+p^2_{\perp}+\delta}
+\sqrt{m_2^2+p^2_{\perp}+\frac{Q^2}{4}+\delta}}\,,
\nonumber \\
&&-(p_i\!-\!p_f)^2=\frac{4(\beta v)^2}{(1\!-\!\beta^2 v^2)^2}
\frac{(2e_p \!+\! e_+ \!+\! e_-)^2}{4(e_+ \!+\! e_-)^2}
\Big((e_+ \!+\! e_-)^2-4p^2_{\parallel} \Big)\,.
\end{eqnarray}
The very last equation can be cast into a form implying the factor multiplying the
quantity $Q$ as in the other cases. The relevant equation is the following one:
\begin{eqnarray}
&&\beta v=\frac{\alpha Q}{\sqrt{4M^2\!+\!\alpha^2 Q^2}}\,,
\nonumber \\
{\rm with}&&\alpha=\frac{M}{\sqrt{2\sqrt{m_2^2\!+\!p^2_{\perp}\!+\!\delta}
\Big(\sqrt{m_2^2\!+\!p^2_{\perp}\!+\!\delta}
+\sqrt{m_2^2\!+\!p^2_{\perp}\!+\!\frac{Q^2}{4}\!+\!\delta}\Big)}}\,.
\label{eq:pfalpha}
\end{eqnarray}
The quantity $\delta$ appearing in the above equation is a small correction 
that vanishes for equal mass constituents. It is defined as follows:
\begin{eqnarray}
4(m_2^2\!+\!p^2_{\perp}\!+\!\delta )
=\frac{(2e_p \!+\! e_+ \!+\! e_-)^2}{4(e_+ \!+\! e_-)^2}
\Big((e_+ \!+\! e_-)^2-4p^2_{\parallel} \Big)\,.
\label{eq:beta1}
\end{eqnarray}
It however depends on $\beta v$, which complicates
the practical derivation of this last quantity. 

The full equation to be solved reads:
\begin{eqnarray}
Q^2\frac{(1\!-\!\beta^2 v^2)^2}{4(\beta v)^2} 
=4(m_2^2\!+\!p^2_{\perp}\!+\!\delta )\,.
\label{eq:beta2}
\end{eqnarray}
It supposes to solve an equation of the 4rth degree in the variable $\beta^2$. 
The correction has been accounted for in the present work. Actually, 
we found simpler to solve the equation by using an iterative process 
(a few iterations were sufficient). Moreover, as the quantity $\delta$ involves
terms depending on  $(m_1^2-m_2^2) $, one could wonder about the appearance 
of $m_2^2 $ instead of $m_1^2 $ or any combination of the masses 
in eq. (\ref{eq:beta1}). The choice made here sounds to provide 
a faster convergence of the iterative process.

Despite differences in expressions, all form factors obtained in different
approaches at $Q^2=0$ can be reduced to a common expression given 
in terms of the internal variable $\vec{k}$:
\begin{equation}
F_1(Q^2=0)= \frac{16\pi^2}{N} \!\int\!  \frac{d\vec{k}}{(2\pi)^3}\; 
\frac{(e_{1k}+e_{2k})}{ 2\,e_{1k}\,e_{2k}}\; 
\tilde{\phi}^2(\vec{k}^2)=\frac{16\pi^2}{N} \!\int\!  
\frac{d\vec{k}}{(2\pi)^3} \;\phi_0^2(\vec{k})
=1 \, .
\label{eq:norm2}
\end{equation}

It was mentioned in the main text that strong effects, before accounting for
constraints related to covariant space-time translations, were due to the
smallness of the pion mass in comparison with the sum of the quark masses.
Looking at the expressions for form factors, it is seen that only 
the instant-form result (Breit frame) does not depend on the mass of the system. 
For the front form with a momentum transfer perpendicular 
to the front orientation ($q^+=0$), the dependence on $M$, through 
the dependence on $E$, can be shown to have no effect by using 
the alternative variable $x=\frac{e_p+p_{\parallel}}{E}$ instead 
of $p_{\parallel}$. For the front form with a momentum transfer parallel 
to the front orientation, a similar trick can be used but only after accounting
for the above mentioned constraints. For the ``point form", the dependence on
$M$ appears through the quantity $\beta v$. By inserting  in the first of eqs.
(\ref{eq:pfalpha}), the value of $\alpha$ obtained from the second one, it is
easily seen that the dependence of $\beta v$ on $M$ cancels. 
The way the dependence on $M$ disappears in this last case differs 
from the one for the front-form case. 


\end{document}